\documentclass{ar-1col}

\usepackage[comma]{natbib}
\usepackage{url}
\jname{Claudia Bank}
\jvol{AA}
\jyear{2022}
\doi{10.1146/((please add article doi))}

\begin{document}

% Page header
\markboth{Bank}{Fitness Landscapes and Epistasis}

% Title
\title{Epistasis and Adaptation on Fitness Landscapes}

%Authors, affiliations address.
\author{Claudia Bank$^{1,2}$
\affil{$^1$Institute of Ecology and Evolution, University of Bern, 3012 Bern, Switzerland; email: claudia.bank@unibe.ch}
\affil{$^2$Swiss Institute of Bioinformatics, 1015 Lausanne, Switzerland}}

%Abstract
\begin{abstract}
Epistasis occurs when the effect of a mutation depends on its carrier's genetic background. Despite increasing evidence that epistasis for fitness is common, its role during evolution is contentious. Fitness landscapes, mappings of genotype or phenotype to fitness, capture the full extent and complexity of epistasis. Fitness landscape theory has shown how epistasis affects the course and the outcome of evolution. Moreover, by measuring the competitive fitness of sets of tens to thousands of connected genotypes, empirical fitness landscapes have shown that epistasis is frequent and depends on the fitness measure, the choice of mutations for the landscape, and the environment in which it was measured. Here, I review fitness landscape theory and experiments and their implications for the role of epistasis in adaptation. I discuss theoretical expectations in the light of empirical fitness landscapes and highlight open challenges and future directions towards integrating theory and data, and incorporating ecological factors.
\end{abstract}

%Keywords, etc.
\maketitle

%Table of Contents
\tableofcontents

% Heading 1
\section{INTRODUCTION}
Determining the relationship between genotype and fitness is a fundamental question in evolutionary biology and beyond. The concept of the fitness landscape quantifies this relationship. It provides an intuitive picture of how selection can direct evolution, iconically visualized by \citet{Wright1932-ds} as a topographic map with peaks and valleys. In a fitness landscape, populations evolve on fitness-increasing paths (via beneficial substitutions) until they reach a fitness peak (the fittest genotype in its mutational neighborhood). A rugged fitness landscape has multiple adaptive peaks representing different, sometimes suboptimal, stable genetic solutions to the challenge of surviving (and thriving) in a given environment. A necessary condition for the existence of more than one adaptive peak is epistasis for fitness, which occurs when one or more mutations interact \citep{Wright1931-fi}. (Throughout this review, I often refer to epistasis for fitness as ``epistasis".) Thus, questions about the shape of the fitness landscape are tightly linked to questions about the prevalence of epistasis.

The role of epistasis in evolution is controversially discussed across the subfields of evolutionary research. Whereas epistasis is considered an important component of reproductive isolation between species when it occurs as hybrid incompatibility \citep{Seehausen2014-rg}, it is usually not considered in studies of population and quantitative genetics. For example, population-genetic selection screens assume that the effect of a selected mutation is independent of the genetic background it occurs on, whereas in quantitative genetics, epistasis tends to appear as a noise term \citep{Walsh2018-ou}. That is because epistasis is considered a transient,  second-order effect that matters only at intermediate allele frequencies \citep{Whitlock1995-zo}. Moreover, incorporating epistasis into evolutionary studies is difficult because of the complexity of multi-locus interactions, resulting in theoretical and statistical challenges \citep{Mackay2013-tp}. Conversely, in molecular and systems biology, epistasis is considered a natural consequence of the structural interaction of mutations within a protein and the interaction of proteins in biological pathways \citep{Domingo2019-vy}. The research area of fitness landscapes bridges evolutionary and systems biology and places epistasis at the heart of its study \citep{Fragata2019-tn}. Fitness landscape theory provides expectations of how epistasis affects evolutionary trajectories, and empirical fitness landscapes provide information on the prevalence of epistasis in nature that feeds back into theory development.

In this review, I take a fitness-landscape view of the role of epistasis in evolution. The first part of the review highlights how epistasis constrains adaptation in two classical but different classes of fitness landscape models, and introduces how  genotype-phenotype or phenotype-fitness maps can contribute to epistasis and its change across environments. The subsequent sections review various empirical fitness landscape studies and evidence for epistasis in evolutionary trajectories that have occurred in nature. I then highlight empirical evidence for environment-dependent epistasis and how this is being integrated into fitness landscape models. The review finishes with open questions and a call for new approaches that incorporate ecological factors into the study of fitness landscapes.
\section{FITNESS LANDSCAPE THEORY HIGHLIGHTS THE ROLE OF EPISTASIS IN EVOLUTION}
The ideal genotype-fitness landscape maps every possible genotype to its carrier's fitness. Even for small genomes, the resulting sequence space is much larger than the number of molecules in the known universe \citep{Szendro2013-pw}. The topological features (e.g., the possibility of fitness ridges between two genotypes) of such a high-dimensional space are difficult to intuit or illustrate visually \citep[see, e.g.,][]{McCandlish2011-ve}. Although no experiment will be able to map the complete fitness landscape of an organism, mathematical models can capture this space and quantify the expected evolutionary patterns for a given shape of a fitness landscape. Fitness landscape theory quantifies  both the invariant properties of the complete fitness landscape (e.g., how many peaks \& valleys? How frequent is epistasis?) and the evolutionary dynamics on the fitness landscape (e.g., which peaks can be reached with which probability? How many steps does it on average take to reach a fitness peak?), and characterizes how they vary with the model parameters (e.g., the strength of epistasis). 

In the following, I introduce two classes of fitness landscape models: probabilistic genotype-fitness landscape models with few parameters, in which epistasis is tunable, represented by the Rough-Mount-Fuji (RMF) model \citep{Aita2000-qw}; and Fisher's Geometric model (FGM; \citealp{Fisher1930-mk}), which creates an epistatic genotype-fitness landscape through its phenotype-fitness mapping. The two approaches highlight different modeling perspectives: whereas probabilistic fitness landscapes directly map genotypes to fitness with a given amount of epistasis, FGM  \citep[as defined by][and onwards]{Martin2007-mq} ignores the possibility that epistasis could occur in the mapping of genotype to phenotype; here, epistasis arises as a consequence of stabilizing selection for a phenotypic optimum. This difference between the RMF and the FGM models highlights a feature inherent to all modeling: the model is greatly influenced by the philosophy of its developer, which affects the results that can be obtained. For example, Fisher considered adaptation to be a process driven by additive substitutions of small phenotypic effects towards a single optimum \citep{Fisher1930-mk}, whereas Wright believed epistasis to be pervasive at the phenotype level and the fitness landscape to be rugged \citep{Wright1932-ds}. Both RMF and FGM-based approaches show how epistasis constrains the course of evolution and predict patterns that were observed in recent empirical studies of fitness landscapes.
\subsection{Probabilistic fitness landscape models feature tunable epistasis}
Probabilistic fitness landscape models allow for the definition of arbitrarily epistatic fitness landscapes with few parameters and therefore lend themselves to mathematical analysis of the consequences of epistasis \citep{Krug2021-dp}. They are agnostic to any mechanistic or molecular reasons for epistasis and are only based on probability distributions. The RMF model \citep{Aita2001-ub,Aita2000-qw} is highlighted here as representative of this class of models; see \citet{Hwang2018-an, De_Visser2014-vr, Ferretti2016-ss} for descriptions of the various other members of this class, including the prominent NK model \citep{Kauffman1989-wb}. In the RMF model, each locus has an additive effect, either taken to be a constant or drawn from a probability distribution, and each genotype is assigned an epistatic deviation from its additive expectation, drawn from a normal distribution with mean 0. 
\begin{figure}[ht]
\includegraphics[width=5in]{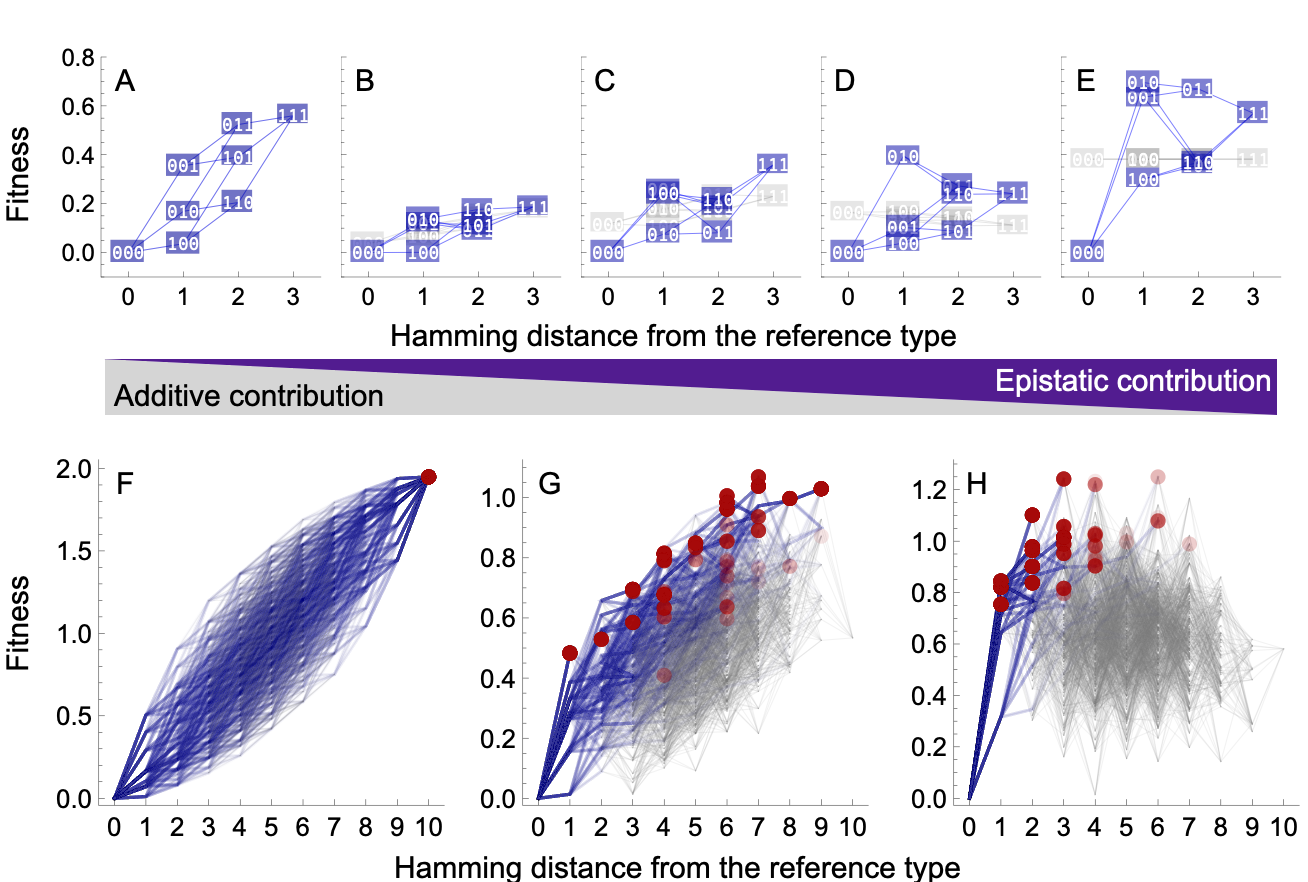}
\caption{Illustration of the Rough-Mount-Fuji (RMF) model. (A)-(E). Examples of three-locus RMF landscapes with increasing epistasis, ranging from additive (panel (A)) to the House-of-Cards model with maximum epistasis (panel (E)). The genotypes are ordered on the x-axis by their Hamming distance (i.e., the number of substitutions distinguishing two genotypes) from the least fit reference genotype 000, whose relative fitness is arbitrarily normalized to 0. Single mutational steps are indicated by blue lines, blue boxes indicate binary genotypes. The underlying additive contribution to the fitness landscapes is highlighted by gray boxes and lines. More fitness landscapes with the same parameters are shown in \textbf{Figure S1}. (F)-(H) Evolutionary trajectories are greatly affected by epistasis. For three landscapes with 10 loci (i.e., $2^{10}=1024$ genotypes) with varying strength of epistasis, 1000 simulations of random adaptive walks were performed, starting at the least fit genotype 000. The underlying fitness landscape showing the fitness of all 1024 genotypes and their connection by single-step substitutions is drawn in gray. The evolutionary paths that were taken are drawn on the fitness landscape in blue, paths that were repeatedly taken are indicated by higher color intensity. Local fitness peaks that were reached are highlighted in red, with color intensity related to the frequency of how often this adaptive peak was reached. See also \textbf{Figure S2}; the simulation code is available as Supplementary Material. In an additive fitness landscape, the single adaptive peak is reached via many different evolutionary trajectories (G). At intermediate epistasis, many of the existing adaptive peaks can be reached but evolutionary trajectories become constrained by epistasis (H). At maximum epistasis, few evolutionary paths remain accessible and only the local fitness landscape is explored by the adaptive walks (I).}
\label{fig1}
\end{figure}
The RMF model interpolates between a fully additive fitness landscape, in which each substitution contributes a fixed amount to its carrier's fitness independent of the rest of the genome (i.e., there is no epistasis), and the so-called House-of-Cards fitness landscape \citep{Kauffman1987-ed}, which maximizes epistasis by assigning a random fitness to each genotype (\textbf{Figure \ref{fig1}}). The difference between the two limiting cases highlights an essential consequence of epistasis: it makes fitness of a genotype less predictable. For example, for a fitness landscape with $L=10$ loci and two alleles per locus, resulting in $2^L=1024$ genotypes, only 10 measurements are required to correctly predict the fitness of all genotypes in an additive fitness landscape, whereas 1024 measurements are necessary to fully determine the House-of-Cards landscape.

Probabilistic fitness landscapes such as the RMF model are an ideal testing ground to study the evolutionary consequences of epistasis when the landscape is tuned from additive to epistatic \citep{Neidhart2014-gy}. In the smoothest, non-epistatic landscape, there is only a single, global fitness peak and any population will eventually climb this peak (\textbf{Figure \ref{fig1}A,F}). At the same time, the evolutionary history of independently evolved populations will tend to be different since any order of substitutions is possible. With maximum epistasis, the RMF landscape is equivalent to the House-of-Cards landscape and displays a large number of fitness peaks, which is on average $2^L/(L+1)$ (\citealp{Kauffman1987-ed}; \textbf{Figure \ref{fig1}E,H}). Overall, epistasis introduces the potential for parallel fitness evolution towards different adapted genotypes (local fitness peaks). However, an increasing number of peaks in the landscape decreases the length of adaptive trajectories and less of the full landscape is explored by evolution. Consequently, well-adapted genotypes are most divergent for intermediate ruggedness of the landscape (\textbf{Figure S2}). Moreover, epistasis tends to create historical contingencies of adaptation. It limits the possible orders of substitutions and thus the number of evolutionary trajectories towards a fitness peak, making evolution from the same starting point more repeatable and seemingly more predictable. (\textbf{Figure \ref{fig1}, Figure S2}). Notably, the regularity of the RMF landscape (and other probabilistic fitness landscape models, \citealp{Hwang2018-an}), in which every genotype has the same statistical properties, is an unrealistic assumption for real interaction networks between mutations or proteins. Nevertheless, the RMF model was shown to recover features of various experimental fitness landscapes \citep{Szendro2013-pw, De_Visser2014-vr, Bank2016-fn}.
\subsection{Epistasis arises from a non-linear phenotype-fitness map in Fisher's Geometric model}
\begin{figure}[ht]
\includegraphics[width=4in]{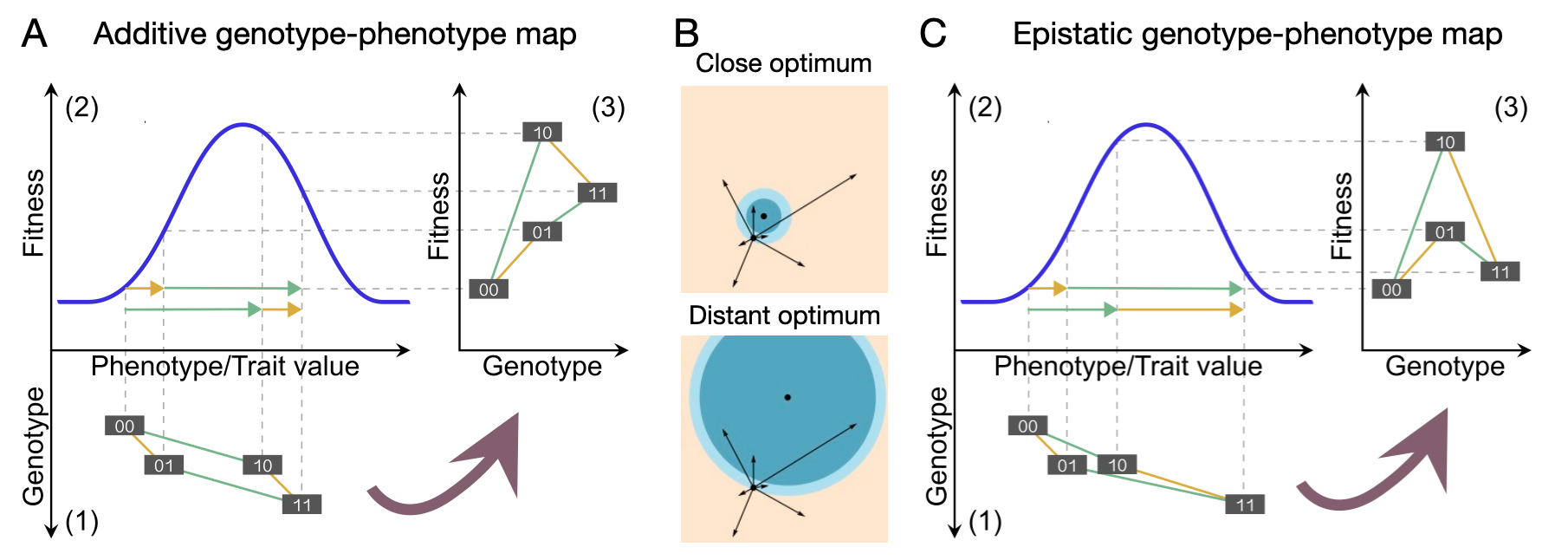}
\caption{(A) In Fisher's Geometric Model, a non-linear phenotype-fitness map introduces epistasis on the genotype-fitness level although mutations are additive on the genotype-phenotype level. The upper left quadrant (2) shows a Gaussian phenotype-fitness relationship (blue curve) in a one-dimensional phenotype space. Two mutations are represented by yellow and green arrows. There is no epistasis on the genotype-phenotype level, i.e., the genotype-phenotype map is additive (lower left quadrant (1); the pairs of yellow and green lines, indicating single-step substitutions, are parallel and of equal length). The Gaussian phenotype-fitness function here results in negative epistasis on the genotype-fitness level (upper right quadrant (3); the pairs of yellow and green lines are at different angles and of different length). Binary genotypes are indicated by gray boxes. (B) In FGM (here visualized in a two-dimensional phenotype space), the same mutations (black arrows) have different fitness effects, determined as the distance gain or loss towards the optimum, when they occur close to (left panel) or far away (right panel) from the optimum (black dot). This occurs when the same mutations are tested on differently well adapted genetic backgrounds in the same environment (resulting in epistasis), or when the same mutations are tested in environments with different phenotypic optima (resulting in genotype-by-environment interaction).  (C) An epistatic genotype-phenotype map (lower left quadrant) and a non-linear phenotype-fitness map with stabilizing selection for an optimum phenotype (upper left quadrant; blue curve) interact to result in an epistatic genotype-fitness relationship (upper right quadrant). Binary genotypes are indicated by gray boxes, connected by single-step substitutions (yellow and green lines).}
\label{fig2}
\end{figure}

Phenotype-fitness models such as Fisher's Geometric Model (FGM; \citealp{Fisher1930-mk}, reviewed in \citealp{Tenaillon2014-ns}) provide a perspective on epistasis complementary to that of probabilistic fitness landscape models. Importantly, under FGM, mutations that are additive on the phenotypic level can be epistatic on the fitness level, because phenotype and fitness are connected by a nonlinear fitness function (\textbf{Figure \ref{fig2}A}; \citealp{Martin2007-mq, Gros2009-kq}). In FGM, fitness is determined by the distance of a phenotype from a single phenotypic optimum in a multidimensional trait space, usually under the assumption of a Gaussian fitness function (i.e., there is stabilizing selection for an optimum phenotype). New mutations appear as jumps in a random direction originating in the current phenotype. A mutation is considered beneficial when it decreases the distance from the phenotypic optimum. Epistasis occurs, for example, when the same mutation appears at different distances from the optimum (\textbf{Figure \ref{fig2}B}). 
FGM turns into a genotype-fitness landscape by introducing an explicit underlying genotype-phenotype map, allowing for the comparison of FGM-based and experimental fitness landscapes \citep{Blanquart2014-jj, Hwang2017-uu}. Interestingly, the single-peaked phenotype-fitness landscape of FGM can give rise to an underlying genotype-fitness landscape with multiple peaks \citep{Hwang2017-uu, Park2020-hn, Srivastava2022-sq}. The amount of epistasis for fitness at the genotype level depends on the distance from the fitness peak: far away from the optimum, the same mutation tested on nearby genetic backgrounds tends to be non-epistatic, whereas strong epistasis can occur when the same mutation finds itself on nearby genetic backgrounds close to the phenotypic optimum. Interestingly and importantly for experimental design, the sampling choice of mutations from the greater fitness landscape influences the shape of the combinatorily complete subset of genotypes built by those mutations: individually beneficial substitutions, measured in a reference genetic background, will tend to result in a more epistatic fitness landscape than substitutions that have occurred subsequently in an adaptive walk \citep{Blanquart2014-jj}. % \citet{Blanquart2016-ti} developed a statistical approach to infer parameters such as the distance from the optimum from empirical fitness landscapes under FGM. Whereas the inferred statistical properties of empirical fitness landscape tended to be compatible with FGM, it could not explain some features of empirical landscapes such as very strong epistasis or large fitness variances.
\subsection{Modeling assumptions and challenges}\label{challenges}
The last subsections showed that in FGM, epistasis arises from a non-linear phenotype-fitness map, whereas RMF and other probabilistic fitness landscape models directly map genotype to fitness without any assumptions about the mapping of genotype to phenotype(s) to fitness. Importantly, epistasis for fitness can arise both through an epistatic genotype-phenotype map and a non-linear phenotype-fitness map, and these levels may interact to yield a (potentially epistatic) genotype-fitness map (\textbf{Figure \ref{fig2}C}). This leads to an open question: how much epistasis for fitness can be attributed to an epistatic genotype-phenotype map, and how much is due to non-linear phenotype-fitness relationships?

Ultimately, the genotype is connected to fitness by multiple layers of multiple phenotypes, and thus the picture discussed here and sketched in \textbf{Figure \ref{fig2}C} is greatly oversimplified. 
%For example, there could be ubiquitous epistasis on the genotype-phenotype level (e.g., for gene expression), which gets diluted towards higher-level phenotypes, and ultimately disappears at the level of fitness. Moreover, the environment may play a major role at determining fitness-relevant phenotypes and tune both the genotype-phenotype and phenotype-fitness maps. 
In an elegant theoretical study, the propagation of epistasis across hierarchical levels in a metabolic network was recently studied by \citet{Kryazhimskiy2021-sv}. They found that negative epistasis cannot be converted into positive epistasis at higher levels, whereas the opposite is possible. Interestingly, their work suggests that epistasis does not get diluted at higher levels but tends to become stronger. % Coming from a mechanistic model, these results support the prediction (and the empirical evidence) that epistasis for fitness is often negative. From an evolutionary point of view, widespread negative epistasis results in ample opportunity for hybrid incompatibilities.

Obtaining general mathematical results about evolution on theoretical fitness landscapes requires additional assumptions about the evolutionary dynamics. Most genotype-fitness landscape models address haploid, non-recombining organisms in the limit of strong selection and weak mutation (i.e., neglecting neutral substitutions and assuming that mutations appear and fix sequentially \citealp{Gillespie1984-nl}; see \citealp{Fragata2019-tn} for a review considering fitness landscape models with neutral substitutions and the consequences of genetic drift). For example, when recombination is included in the study of adaptive walks on fitness landscapes, the connectedness of genotypes in the fitness landscape changes because recombination produces new combinations of (multiple) segregating alleles \citep{Moradigaravand2012-kw, Nowak2014-wy}. For rugged RMF landscapes, recombination creates an initial advantage to recombining populations that is later lost during adaptation because recombination dilutes genotypes necessary for crossing fitness valleys. 
%Specifically, recombination initially creates extra diversity that can be selected upon, but later breaks up beneficial genotypes when the path to the next fitness peak requires passing through a valley spanning at least one substitution. 
Thus, theory predicts that recombining populations adapting on an epistatic fitness landscape can get stalled at local optima even more severely than non-recombining populations \citep{Nowak2014-wy}.

Studying diverse and recombining populations on fitness landscapes creates conceptual and technical questions. In a diploid fitness landscape, dominance has to be accounted for with respect to the direct effect of individual mutations and their epistasis because it affects both the duration for which epistatic alleles segregate and their ultimate fate in a population \citep[see, e.g.][]{Turelli2000-ez, Blanckaert2018-ku}. For example, there is currently no commonly used biologically reasonable and parameter-light definition of the dominance of epistasis, i.e., of the effect of combinations of one or two copies of sets of epistatic alleles.

Finally, population dynamics considering multiple coexisting genotypes are difficult to characterize mathematically and to simulate computationally. If a starting population with standing genetic variation is to be considered, how should this population be distributed in the fitness landscape? Finding simple, computationally feasible, and yet biologically relevant assumptions is a major challenge underlying theoretical studies of fitness landscapes. 
% Moreover, the consideration of some of the above-mentioned factors and linking this theoretical work to observed patterns in nature would bring the field of fitness landscape theory to greater attention among empirical biologists. %\begin{textbox}[h]\section{GENERAL THEORETICAL PREDICTIONS}
%\begin{enumerate}
%\item Epistasis reduces the number of available adaptive paths.
%\item There is an excess of positive epistasis at fitness peaks.
%\item Diminishing-returns epistasis occurs as a population adapts in a fitness landscape.
%\item A non-linear phenotype-fitness map can convert positive or no epistasis on the genotype-phenotype level into negative epistasis on the genotype-fitness level. 
%\item The sampling choice of mutations for an experimental fitness landscapes affects its shape.
%\end{enumerate}
%\subsection{Sidebar Second-Level Heading}
%More text goes here.\subsubsection{Sidebar third-level heading}
%Text goes here.
%\end{textbox}
\section{EPISTASIS IS COMMON IN EXPERIMENTAL FITNESS LANDSCAPES}
Recent advances in technology and experimental approaches (including CRISPR and site-directed mutagenesis to create sets of specific mutations \citealp[e.g.,][]{Karageorgi2019-os,Hietpas2012-iw}, and deep mutational scanning, in potential combination with barcoding approaches, to measure phenotypes or fitness of thousands of genotypes in parallel \citealp[e.g.,][]{Fowler2014-qr,Cote-Hammarlof2021-ow,Johnson2019-ya}) have enabled the empirical assessment of increasingly large empirical fitness landscapes and the subsequent evaluation of fitness landscape theory. When selection is strong (e.g., in large microbial populations), and if these findings translate to natural environments, empirically tuned fitness landscape models promise to be useful for predicting evolution, which is relevant in human medicine (e.g., antimicrobial resistance or cancer evolution) and with respect to climate change (e.g., potential for evolutionary rescue under the climate crisis) \citep{De_Visser2014-vr, Lassig2017-vw, Wortel2021-ie}. 
\subsection{Engineered fitness landscapes measured in the laboratory}
% Two main challenges were overcome to make the study of large experimental fitness landscapes possible. Firstly, desired/known and defined genotype and combinations of mutations had to be created artificially. When a combinatorially complete fitness landscape is to be assessed, it is necessary to experimentally create all possible combinations of all considered mutations. This challenge was overcome by the development of sophisticated reverse genetics methods such as CRISPR (e.g., \citealp{Karageorgi2019-os}) and site-directed mutagenesis (e.g., \cite{Hietpas2012-iw}). Secondly, a fitness measurement had to be obtained for each genotype, ideally based on competitive fitness. This challenge was overcome by developing a combination of bulk competitions and deep sequencing (yielding the change in relative frequencies of all genotypes over time), the main approach of which was termed deep mutational scanning \citep{Fowler2014-qr}. Recently, the addition of barcoded pieces of sequence has made this approach even more powerful and accurate (e.g., \citealp{Cote-Hammarlof2021-ow, Johnson2019-ya}).
As an intermediate between theoretical fitness landscapes and data from nature, experimental fitness landscapes are appealing for several reasons. Firstly, depending on the design of the mutant library they may contain all genotypes of interest, not only those which are segregating naturally; thus, deleterious or weakly beneficial genotypes can be measured, and inference of the selection coefficient (i.e., the relative competitive advantage of a genotype) can be performed for each genotype, resulting in a measurement of the (potentially combinatorially complete) fitness landscape. Secondly, the lab environment is controlled such that all genotypes grow or evolve in a constant, known, and similar environment. Finally, samples can be resequenced to check for secondary mutations, which are anyways minimized in the setup of deep mutational scanning, where populations are observed for only a few generations. Together, these features result in elegant and reproducible fitness data.

Many fitness landscape studies come from systems and molecular biology, where epistasis is considered a natural consequence of interactions within and between genes (reviewed in \citealp{Domingo2019-vy, Phillips2008-tr, Costanzo2019-rb, Athavale2014-bk}). Because intra- and intergenic interactions intuitively arise from the physical organisation of proteins and the connection of genes in biological pathways, the question is not whether there is epistasis but how much, and which models of epistasis are most suitable. %(but note that for evolutionary consequences of this epistasis, the epistatic alleles would have to segregate in evolving populations). 
Among a multitude of studies that screened hundreds or thousands of engineered genotypes for (usually pairwise) epistasis, most have focused on measurements of phenotypes that may be fitness-related but do not directly translate into competitive fitness. Across these studies, pairwise epistasis was found to be common and there is increasing evidence of higher-order epistasis \citep{Domingo2019-vy}. Moreover, an increasing number of studies have followed evolutionary trajectories in the laboratory and dissected epistatic effects along those trajectories \citep[e.g.][]{Kryazhimskiy2014-yn, Johnson2019-ya}. This review focuses on engineered slices of fitness landscapes built from combinations of more than two mutational steps and with an emphasis on studies measuring competitive fitness. 
\subsection{Experimental design influences the shape of the fitness landscape}
\subsubsection{Which level of biological organization was considered?}
Three main components distinguish experimental fitness landscapes studies that may explain some of the observed differences between fitness landscapes and that have direct implications for the connection of models and data.
Firstly, studies differ by the considered level of biological organization. For example, a landmark study of all 32 combinations of five mutations in the beta-lactamase gene in {\em E. coli} that together led to a $10^5$-fold increase in resistance to the antibiotic cefotaxime reported that only 18 out of 120 possible evolutionary trajectories from the reference genotype to the resistant genotype in the fitness landscape were accessible, i.e., they had monotonically increasing fitness steps \citep{Weinreich2006-dr}. In contrast, when \citet{Khan2011-lr} performed competition experiments of the combinations of the first five adaptive substitutions that had been fixed during laboratory evolution of {\em E. coli} in a glucose-limited minimal medium, the majority (86) of the 120 possible evolutionary trajectories to the fittest genotype were accessible. Here, \citet{Khan2011-lr} studied substitutions in different genes, whereas \citet{Weinreich2006-dr} studied a set of point mutations in the same gene.

On average, one might expect more epistasis between random alleles within the same gene than between random alleles between genes, because the physical structure of a protein leads to locally interacting residues and fine balances in the composition of the protein sequence. For example, \citet{Bank2015-kh} reported extensive epistasis within the heat-shock protein Hsp90 in {\em S. cerevisiae}, where almost 50\% of $\approx$1000 interactions of pairs of point mutations in a 9-amino-acid region of the protein showed negative epistasis for competitive fitness, compared with $\approx$3\% of significant fitness interactions (based on colony size as measure of fitness) of double gene knockouts in the same species \citep{Costanzo2010-cn}. These numbers have to be interpreted carefully: in addition to the fitness proxy, the measurement accuracy between these studies differed greatly. Moreover, an enrichment in epistasis is expected between genes located in the same pathway than between random genes \citep{Hurst2004-vn}. A systematic metaanalysis or a dedicated experimental study could better evaluate these expectations in the future.   

Theoretically, knowledge of the structure of a protein, the interactions of genes within the protein, and the phenotypes important for fitness in a given environment could be used to develop predictive models of epistasis. \citet{Kemble2020-kb} studied the competitive fitness of 1369 genotypes in {\em E. coli} carrying mutations in two adjacent metabolic genes in the same pathway. The shape of their double-mutant fitness landscape strongly depended on the induced amount of gene expression in the pathway (which was tuned experimentally here and could represent different environmental pressures). This study stands out because the authors developed a specific model of the metabolic pathway that captured the severe changes in fitness effects and epistasis occurring across the different expression levels at impressive accuracy. Here, the model allowed for an accurate quantitative analysis of the specific scenario; at the same time its specificity prevents straightforward extrapolation of the results to other pathways or higher-level questions such as the role of epistasis in evolution. Taking this to the organismal level, systems biologists have developed complex genome-scale metabolic models that predict phenotypes across environments \citep{King2016-ld, Bordbar2014-cg}. It would be interesting to see how such predictions relate to competitive fitness of bacterial strains in the lab or observed sequence variation in nature.

In general, the arguments above indicate that epistasis is expected to be most prevalent at short genomic distances, when substitutions within genes or in genes within the same pathway interact. At the same time, the evolutionary relevance of epistasis at short physical distances in the genome is unclear. Even if variation was present at physically close loci, it would rarely be exposed to epistatic selection due to the low recombination probability between closely linked loci and the low probability that the epistatic mutations occur in the same genotype. Evolutionary models of complex fitness landscapes incorporating recombination could shed light on how this dichotomy affects the expected distribution and role of epistatic variation in natural populations.
\subsubsection{Which measure of fitness was used?}
Secondly, many studies have measured phenotypes other than competitive fitness such as monoculture growth rate (e.g., \citealp{Pinheiro2021-qq}), fluorescence (e.g., \citealp{Sarkisyan2016-sj}), antibody binding (e.g., \citealp{Adams2019-vg}), minimal inhibitory concentration in the presence of a drug (e.g., \citealp{Weinreich2006-dr}, ribozyme activity (e.g., \citealp{Bendixsen2021-iz}), or transcription factor binding (e.g., \citealp{Aguilar-Rodriguez2017-gz}). Often, there is a clear relationship between the measured phenotype and reproductive success (e.g., if a protein is essential, its stability is necessary for survival of the individual). However, how the fitness-related phenotype maps to competitive fitness is generally unknown, and even bacterial growth rates in monoculture, which are often treated as equivalent to competitive fitness, are correlated with but do not translate directly into competitive fitness \citep{Chevin2011-zu, Concepcion-Acevedo2015-nn}. As shown in Figure \ref{fig2} and discussed in Section \ref{challenges}, epistasis on the genotype-phenotype level can differ greatly from epistasis for fitness, which implies that genotype-phenotype landscapes may often be bad predictors of the true fitness landscape.  
% Some mappings between lower-level traits and the measured phenotypes are well understood and can explain the ubiquitously observed negative epistasis \citep{Domingo2019-vy}.
\subsubsection{How were the genotypes for the fitness landscape chosen?}
Finally, the choice of the mutations for an experimental fitness landscape affects its shape \citep{Blanquart2014-jj}. \citet{Khan2011-lr} combined five substitutions that had occurred and fixed together in adaptive walk during laboratory evolution. Due to this choice, the 5-step genotype was by definition highly fit, and at least one evolutionary trajectory between the reference and the endpoint had to be accessible. Using a different sampling strategy, \citet{Bank2016-fn} studied the complete fitness landscape of 640 genotypes in {\em S. cerevisiae} of point mutations (all within a 9-amino-acid region in the heat-shock protein Hsp90) that had not been previously found together and were individually beneficial to slightly deleterious. The resulting fitness landscape of up to 6 combined mutations showed both strong positive epistasis (at the local and global peaks) and strong negative epistasis, with many strongly deleterious combinations of individually beneficial mutations. Interestingly, none of the high-fitness genotypes observed in this fitness landscape were found in natural sequence data. The observed results were compatible with theoretical predictions: mutations that occurred together in an adaptive walk were less epistatic than random mutations \citep{Blanquart2014-jj}, there was an excess of positive epistasis at local peaks \citep{Greene2014-uu}, and negative (diminishing-returns) epistasis was common (discussed below in Section \ref{dimret}).

With increasingly large experimentally measured fitness landscapes, it has become possible to consider different levels of mutation choice in the same landscape. \citet{Pokusaeva2019-nl} measured competitive fitness resulting in 12 fitness landscapes in {\em S. cerevisiae} spanning a total of $>$4 million genotypes, including the majority of extant amino-acid substitutions of the his3 protein across 21 yeast species. Their fitness landscapes were generally rugged with 85\% of substitutions being genetic-background dependent, and the evolutionary trajectory was constrained by epistasis. Although commonly observed and theoretically predicted patterns such as negative epistasis away from fitness peaks or ridges were confirmed in the large data set, simple models of sigmoidal phenotype-fitness relationships were not sufficient to describe the data; only a complex multi-level model captured the observed epistatic interactions.
\subsection{Pervasive diminishing-returns epistasis points to ubiquitous interactions on the genotype-phenotype level}\label{dimret}
Although the extent of inferred epistasis in experimental fitness landscapes differed between study system, measure of fitness, and environment, many studies shared the observation of negative (diminishing-returns) epistasis, where the effect of the same beneficial mutation was smaller when it was measured in a higher-fitness genetic background (see \citealp{De_Visser2014-vr} for an earlier review). Diminishing-returns epistasis results in a gradual slow-down of fitness evolution and is predicted across various fitness landscape models when the population approaches a fitness peak \citep{Martin2007-mq, Lyons2020-lf, Draghi2013-hl, Greene2014-uu}. Diminishing-returns epistasis is  predicted generally during adaptation on a fitness landscape when there is widespread ``idiosyncratic'' epistasis, corresponding to epistasis on the genotype-phenotype level in \textbf{Figure \ref{fig2}C}; \citet{Lyons2020-lf, Reddy2021-ec}). Models of idiosyncratic epistasis can, for example, explain he fitness trajectories obtained from a long-term evolution experiment in {\em E. coli}, in which even after 50000 generations of evolution no fitness peak was reached \citep{Wiser2013-uj}. Interestingly, the epistasis that occurs when an additive trait is translated into fitness through a non-linear phenotype-fitness map (also termed ``nonspecific epistasis"; \citealp{Domingo2019-vy}; see for example the genotype-phenotype-fitness map of FGM in \textbf{Figure \ref{fig2}A}) is not sufficient to guarantee persistent diminishing-returns epistasis \citep{Reddy2021-ec}. Thus, theory indicates that (potentially ubiquitous) epistasis on the genotype-phenotype level may contribute to observed evolutionary trajectories. 
\section{EPISTASIS AND FITNESS LANDSCAPES IN THE WILD}
In an evolving population, epistasis is a second-order effect because it only becomes visible when recombination combines two (or more) epistatic alleles, or when mutation rates are high enough to create epistatic double mutants. Therefore, epistasis is generally believed to be negligible during adaptation on short time scales as compared to other evolutionary forces such as directional selection and genetic drift. However, the widely observed pattern of diminishing-returns epistasis, for example, can lead to a change in the selection gradient over time and is therefore likely to lead to wrong predictions of how quantitative traits change over time \citep{Hansen2021-yl}. Moreover, epistasis creates detectable contingencies between substitutions in distantly related species that have been used to infer interactions between proteins (e.g., \citealp{Morcos2011-di, Marmier2019-db}). The prevalence of epistasis in experimental studies, together with its observed footprint in phylogenies, calls for a reevaluation of how epistasis affects evolutionary inference. 
\subsection{Experimental evidence of epistatic adaptive trajectories in the wild}
Retracing adaptive steps during evolution in the wild is becoming feasible with the advance of experimental methods, especially in organisms that lend themselves to experimental manipulation in the laboratory. However, few studies to date have explicitly screened evolutionary trajectories for the presence of epistasis for fitness (see \citealp[]{Storz2018-en} for a review of studies of genotype-phenotype epistasis). \citet{Pokusaeva2019-nl}, introduced above, built intragenic fitness landscapes including extant variation across various yeast species. They found ubiquitus and often higher-order epistasis. The epistasis resulted in fitness ridges such that fit genotypes tended to be located near extant variation in the sequence space. Thus, most likely epistasis played a role during evolution by defining the accessible sequence space, including potentially neutrally accumulated divergence.

In a study of a protein under strong selection, \citet{Gong2013-wm,Gong2014-zb} dissected the epistatic evolutionary trajectory of the human influenza nucleoprotein (NP) between 1968 and 2007. They used reverse genetics approaches to create genotypes carrying individual substitutions that had occurred during the evolutionary trajectory on the original genetic background from 1968, and used the growth rate of these genotypes in cell cultures as measure of their fitness. \citet{Gong2013-wm} found that several of the substitutions that had fixed during the evolutionary history of NP were individually deleterious in the background of the original sequence, but enabled the virus to subsequently accumulate beneficial immune-system evading substitutions. Interestingly, no epistatic trajectories were observed in NP in swine influenza \citep{Gong2014-zb}.  Whereas humans tend to get infected by influenza multiple times, swine are rarely infected more than once, which leads to much stronger selection for immune invasion in human influenza than in swine influenza. The results from \citet{Gong2013-wm,Gong2014-zb} indicate that strong selection was a prerequisite for observing epistasis in the evolutionary trajectory. Whereas laboratory studies in bacteria support that more strongly selected mutations have stronger interactions \citep[e.g.][]{Schenk2013-tu}, it remains unknown whether epistatic trajectories in nature are enriched for individually strongly selected substitutions. Whether this pattern is predicted by theory depends on the considered model scenario. For example, FGM predicts strong epistasis between strongly selected substitutions when the optimum is close, but not when it is far away. 

One example of the dissection of an evolutionary trajectory in complex organisms comes from \citet{Karageorgi2019-os}, who studied a convergent evolutionary trajectory of the adaptation of multiple insects including monarch butterflies to cardiac glycoside toxins. They overcame the challenges of retracing evolution by using CRISPR-Cas9 based genome editing, which allowed them to create and test combinations of the target mutations introduced into {\em Drosophila melanogaster}. They found strong positive epistasis for toxin tolerance between the three substitutions that had been repeatedly observed during evolution of toxin resistance. In addition, the order at which substitutions could accumulate was found to be constrained by epistasis \citep{Karageorgi2019-os}. Positive epistasis at fitness peaks is a general theoretical prediction that holds independent of the fitness landscape model \citep{Greene2014-uu}.
\subsection{Epistasis is commonly observed in form of hybrid incompatibility}
In the research area of speciation, epistasis plays a major role when it manifests itself as hybrid incompatibility (reviewed in \citealp{Seehausen2014-rg, Coughlan2020-ik}). The classical model that explains speciation via the evolution of postzygotic hybrid incompatibilities in geographically isolated populations is the (Bateson-)Dobzhansky-Muller model. In this model, allopatrically diverging populations accumulate substitutions over time. Any combinations of substitutions not tested by natural selection because they never segregated in the same population have the potential to be incompatible, leading to so-called Dobzhanzky-Muller incompatibilities (DMIs). As pointed out by \citet{Gavrilets1997-ed}, the resulting fitness landscape contains ridges of connected high-fitness genotypes and valleys of incompatible genotypes reflecting negative epistasis between the outermost genotypes on the ridge. Thus, a defining element of a classical DMI is that only one of the recombinant offspring of hybrids suffers from low fitness, whereas the broad definition of hybrid incompatibility includes any combination of alleles reducing fitness in hybrids \citep{Seehausen2014-rg}. 

Hybrid incompatibilities and DMIs between species have been mapped in laboratory crosses and more recently using large genomic screens, often revealing hundreds of pairwise incompatibilities between species pairs. For example, a genetic map of incompatibilities between the two fly species {\em D. melanogaster} and {\em D. simulans} resulted in an estimate of $>$100 hybrid-lethal incompatibility regions \citep{Presgraves2003-yl}. Similarly, between two hybridizing swordtail fish in Mexico, genomic analyses of hybrid genomes indicated $>$100 incompatibility loci \citep{Schumer2014-da}. Also within species, segregating epistatic interactions may be common: \citet{Corbett-Detig2013-rr} found various pairwise epistatic interactions when screening the global genetic variation within {\em D. melanogaster}. 

Measuring a complex fitness landscape in the wild that encompasses more than pairwise interactions is difficult for various reasons. For example, diploidy creates a challenge for defining fitness, it is often difficult to obtain a sufficiently large sample to have a good representation of genotypes, fitness measurements are even less straightforward than in the lab, background variation might affect measurements, or experimental manipulation and creation of fitness variation can be impossible. Hybrid populations circumvent several of these challenges by being natural sources of genetic and fitness variation. Few studies to date have attempted to leverage this opportunity to map multi-locus fitness landscapes in the wild. In a landmark study, \citet{Martin2013-jz} mapped the phenotype-fitness landscape of hybrids of three recently diverged pupfish ecotypes in field enclosures. Based on measures of growth and survival, they found a multi-peaked phenotype-fitness landscape isolating molluscivores from generalists with no evidence of frequency-dependent selection \citep{Martin2013-jz, Martin2020-ws, Martin2016-vp}. In a follow-up study, \citet{Patton2021-mp} recently studied the genotype-fitness landscape underlying the observed fitness differences between specialist and generalist phenotypes. 
% Since the observed genome-wide variation was too large to create a comprehensive fitness landscape, they compared the fitness sub-landscapes of samples of 5 loci from different sources of variation (de-novo, introgressed, and standing variation) and found the specialist peak to be more accessible through de-novo and introgressed variation than through standing genetic variation. It remains to be tested whether the conclusions reflect different ruggedness of the fitness landscape between the different subsets of loci, or a greater contribution of genetic drift to the standing genetic variation. The development of methods to extract complex fitness relationships from hybrid variation, as pioneered in \citet{Patton2021-mp}, is a promising step towards quantifying the prevalence of epistasis in the wild.
\subsection{The contentious role of epistasis during speciation}
Epistasis enables multiple genetic solutions to adaptation, represented by multiple peaks in the fitness landscape. Consequently, epistasis is thought to favor allopatric divergence. However, it is an open question how much epistatic alleles contribute to speciation early during divergence or in the presence of gene flow, and how (much) epistatic variation is accumulated and maintained within populations. \citet{Barton1986-bi} argued that hybrid incompatibilities tend to be weak barriers to gene flow. Evolutionary theory has predicted that, because of the specific shape of their fitness landscape, epistatic alleles in a DMI cannot be maintained polymorphic in a population and that their accumulation in the presence of gene flow can only be a by-product, rather than a driver, of evolution \citep{Gavrilets1997-ed, Bank2012-ao}. However, more complex epistatic interactions might behave differently. \citet{Blanckaert2020-yb} recently showed that adding a third locus to the fitness landscape of a hybrid incompatibility allows for the evolution of complete reproductive isolation in the presence of gene flow (see \citealp{Paixao2015-zo} for a network-based approach). In a series of theoretical works, \citet{Fraisse2016-pp, Simon2018-qt, Schneemann2020-kf} have studied the evolution of hybrid incompatibility using FGM, recovering many of the predictions of previous theoretical models and patterns observed in empirical data. Furthermore, they developed indices characterizing the respective influence of the environment and genetics on hybrid fitness \citep{Schneemann2020-kf}.

Taking a fitness-landscape view of speciation models in which higher-order interactions are possible is challenging due to the complexity of the models (and the potential genomic inference of such interactions), but may be an opportunity to resolve the dichotomy between the frequency at which hybrid incompatibilities are observed and the theoretically ascribed minor role of epistatic loci early during divergence \citep{Seehausen2014-rg}. Given the potential overrepresentation of epistatic loci at short physical distances in the genome (e.g., within proteins, or regulatory interactions), it will be important to address the role of genetic architecture and recombination distances in determining the evolutionary dynamics of epistatic loci (see, e.g., \citealp{Blanckaert2018-ku}). Here, probabilistic fitness landscape models can be a tool to quantify how the amount of epistasis in a fitness landscape affects speciation. For example, with increasing ruggedness, the number of fitness peaks increases and with it the potential for populations to climb different fitness peaks. At the same time these fitness peaks will be located close to each other in sequence space, which should result in a smaller barrier to gene flow when the diverged populations hybridize (see also \textbf{Figure S2}). Thus, an intuitive prediction would be that intermediate epistasis maximizes the potential for stable divergence. 
\section{ENVIRONMENT-DEPENDENT EPISTASIS: MODELS AND EMPIRICAL EVIDENCE}
\subsection{Empirical evidence for environment-dependent epistasis}
\begin{figure}[ht]
\includegraphics[width=5in]{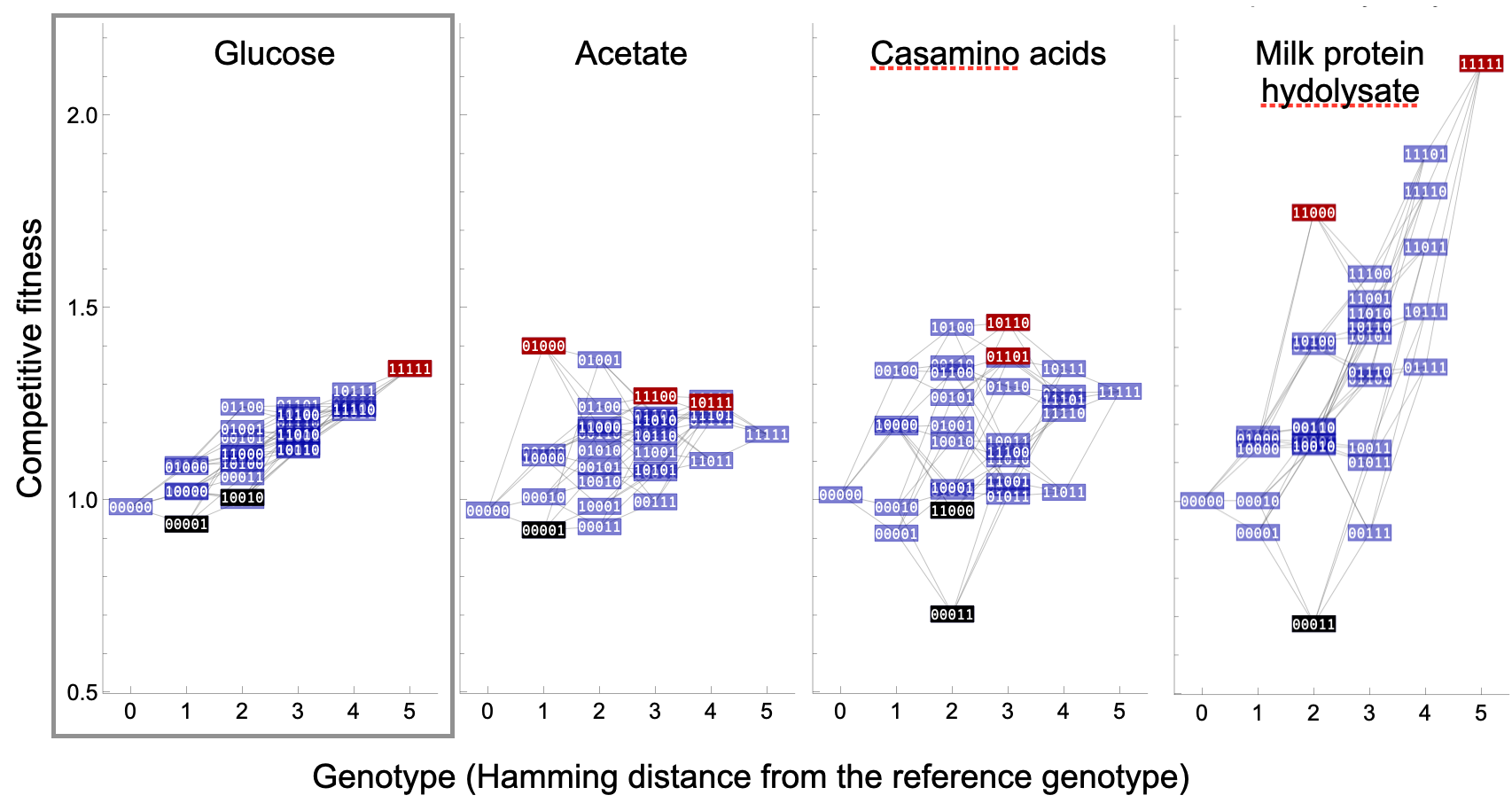}
\caption{Environment-dependent fitness landscapes from \citet{Hall2019-pp}. Each fitness landscape consists of the same $2^5=32$ genotypes, initially studied in \citet{Khan2011-lr}. The five substitutions that were combined in this landscapes had been the first steps of an adaptive walk in a laboratory evolution experiment in a glucose environment. Their competitive fitness was then measured in the original environment (gray box) and 7 additional environments (3 of which are shown here). In each fitness landscape, single mutational steps are indicated by gray lines, blue boxes indicate genotypes, and local fitness peaks and sinks are highlighted in red and black, respectively. The fitness landscapes are vastly different between environments, differing in absolute effect sizes, location and number of peaks and valleys, and accessible evolutionary paths. The fitness landscape in the original evolution environment is the smoothest.}
\label{fig3}
\end{figure}

Despite great advances in the study of both theoretical and empirical fitness landscapes and epistasis, little is known about how fitness landscapes and epistasis change across environments. Whereas experimental screens have reported several cases of ubiquitous genotype-by-environment interactions for fitness (GxE; also referred to as antagonistic pleiotropy and costs of adaptation; e.g., \citealp{Bank2014-xw, Kinsler2020-rz, Flynn2020-si}), the presence and magnitude of environment-dependent epistasis (or genotype-by-genotype-by-environment interactions; GxGxE) for fitness has received little attention to date. The few existing examples have reported striking variation in epistasis between environments. When \citet{Hall2019-pp} competed the 32 {\em E.coli} genotypes from \citet{Khan2011-lr} across eight different environments, the resulting fitness landscapes looked vastly different between environments; epistasis was shown to be both strong and environment-specific (\textbf{Figure \ref{fig3}}; see also \citealp{Flynn2013-gf}). In an experimental study of yeast that had evolved in the presence of a fungicide, epistatic interactions between two point mutations could change greatly dependent on the concentration of the drug \citep{Ono2017-cz}. \citet{Gorter2018-zq} showed that the evolution of yeast populations in the presence of increasing concentrations of nickel followed the evolutionary trajectory predicted by the changes in the fitness landscape along the dosage gradient. \citet{Lindsey2013-fd} showed that environment-dependent epistasis prohibited evolutionary rescue of {\em E. coli} populations under high rates of environmental change. In a high-throughput study of a yeast tRNA fitness landscape across four environments, \citet{Li2018-gs} also reported widespread environment-dependence of epistasis. Finally, \citet{Bakerlee2022-bc} recently reported large environmental variation with respect to both single-mutation effects and pairwise epistasis of a fitness landscape spanning 10 mutations in different genes in yeast, measured in six environments. Based on their analysis the authors argued that epistasis in their fitness landscapes arose from specific gene interactions rather than a global, non-specific, non-linear phenotype-fitness relationship.
Environment-dependent epistasis was also observed in studies of insects and fish. \citet{Nosil2020-nd} mapped the fitness landscape related to cryptic coloration seen in stick insects using a field experiment. Here, the fitness proxy was survival during a two-day exposure to predators in the wild. 
% Interestingly, the results pointed to a change in the phenotype-fitness map between environments as the reason for the observed environment-dependent epistasis. 
In a subset of F2 offspring from a sympatric pair of threespine stickleback populations that otherwise showed little evidence of epistatic interactions, \citet{Arnegard2014-pu} identified an environment-dependent functional mismatch of oral jaw traits contributing to a reduction in feeding performance. Interestingly, both studies pointed to a major role of environment-dependent changes in phenotype-fitness relationships in creating and altering epistasis.
\subsection{Theoretical approaches to environment-dependent epistasis}
Most fitness landscape theory to date has addressed a single, static environment, and thus a static fitness landscape. %However, real environments in nature fluctuate or change, which may have great effects on the evolutionary dynamics and the fitness landscape ``perceived" by a population over evolutionary relevant time scales. 
However, the study of evolution in changing environments is important both from a basic science point of view and with respect to the response of populations to anthropogenic change and the climate crisis. %Moreover, as argued above, many empirical studies have shown genotype-by-environment interactions to be common, which begs the question whether and how epistasis changes across environments. 
Fluctuating phenotypic optima turn a fitness landscape into a ``seascape" \citep{Mustonen2009-cz}. Although the concept of the fitness seascape is gaining increasing interest from the scientific community, little work has addressed how to extend the existing fitness landscape theory to account for changing environments and how this affects epistasis and its consequences. 

Fisher's Geometric Model is the most extensively studied class of fitness landscape models with respect to changing environments \citep{Harmand2017-ic, Martin2015-qz, Martin2006-ad, Zhang2012-xm, Matuszewski2014-ot, Schneemann2020-kf}. Different environments can be incorporated in FGM by assuming a change in the location of the phenotypic optimum (\citealp{Martin2015-qz, Martin2006-ad}, see also Figure \ref{fig2}B). Under this assumption, it is possible to study environmental gradients as well as completely different environments, and to develop expectations of the amount of genotype-by-environment interactions. Using experimental data from {\em E. coli}, evolved at different concentrations of an antibiotic drug, \citet{Harmand2017-ic} could predict the costs of antibiotic resistance across environmental gradients with a generalized FGM. From both a theoretical and experimental point of view, an exciting next step would be to extract the how the underlying genotype-fitness landscape changed across environments.

Several studies of dynamic genotype-fitness landscape models have addressed how populations adapt to fluctuating environments, with a focus on the response to the time-scale at which fluctuations occur and whether evolution results in the repeated switching between specialists or the evolution of generalists \citep{Sachdeva2020-lm, Wang2019-gb, Trubenova2019-xq}. Moreover, \citet{Agarwala2019-mj} showed that in a slowly changing environment, epistasis promotes evolutionary dynamics that switch between short phases of rapid adaptation and long periods of stasis in which the population waits for new adaptive paths to open up. 
\subsection{Environment-dependent epistasis could leave a footprint of ``ecological history" in genomes}
Recent theoretical work by \citet{Das2020-xt} has shown how simple genotype-phenotype relationships can result in environment-dependent epistasis. Their model specifically described the genotype-fitness landscapes along an environmental gradient of antibiotic concentration, based on Hill functions that are commonly used to describe the S-shaped relationship between bacterial growth rate and antibiotic concentration (so-called pharmacodynamics \citep{Goutelle2008-yt}). \citet{Das2020-xt} assumed that every genotype has a different S-shaped fitness curve, and that there is a trade-off between growing well in the absence and the presence of the antibiotic; i.e. resistant genotypes must have a cost of resistance. Their model was in excellent agreement with a previously described experimental fitness landscape from {\em E. coli} \citep{Marcusson2009-nf}. Interestingly,  although \citet{Das2020-xt} observed little epistasis for the two main phenotypic parameters of the model, the drug-free growth rate of genotypes and their IC50 (the concentration of the drug reducing the growth rate by 50\%), epistasis was common in the composite fitness measure combining these two parameters and variable along the dose gradient. Epistasis was greatest at intermediate concentrations of the antibiotic and the fitness ranking of the genotypes changed most rapidly at intermediate drug concentrations. The authors concluded that small changes in (intermediate) dosage might have large consequences for evolution, because new evolutionary routes may open up or close.

Extending their theoretical analysis of the model, \citet{Das2021-kj} found that a genotype can carry information on its ``ecological history"; in other words, the current genotype is informative of the previous environments. This scenario arises in fitness landscapes in which epistatic relationships change along an environmental gradient, such that evolutionary trajectories are contingent on the previously experienced environment. Firstly, some genotypes were only accessible at low doses of an antibiotic drug, and would therefore never be observed if the population experienced high doses of the drug in its recent history. Secondly, there was on average limited reversibility of evolutionary paths, i.e., if the environmental change was reversed the population did not tend to take the reverse evolutionary path. Notably, the obtained analytical results are based on specific models and approaches inspired by theoretical physics. It will be interesting to see if similar results can be obtained from other models incorporating environmental gradients, such as FGM.
\section{CONNECTIONS, DISCREPANCIES, AND FUTURE CHALLENGES}
\subsection{Ecological factors may smoothen the fitness landscape that matters for evolutionary dynamics in the wild}
%Whereas there is an increasing number of studies that have quantified epistasis on a large scale, these often come from laboratory studies in artificial and single environments. In the few studies across environments not only the fitness effects of mutations, but also epistasis, tended to change greatly across environments. Thus, u
Under the assumption of environmental fluctuations in natural environments, evolution may not only be constrained by epistasis, but this epistasis may also be variable across environments, further complicating the study of selective forces that acted during evolutionary history. So far, few models can accommodate the change of epistasis across environments, and there is little knowledge about whether these effects could ``cancel out" integrated over evolutionary history, maybe resulting in, on average, neutral-like evolutionary dynamics. 

At the same time, selection coefficients inferred from empirical fitness landscape studies tend to be much larger than those estimated via inference of the distribution of fitness effects from polymorphism data (reviewed in \citealp{Tataru2020-vv}). This could be due to the complexity of natural environments as compared to experimental settings. Over evolutionarily relevant time scales (i.e., those reflecting the polymorphism used for the above-mentioned inference methods), the experienced environment of a population has likely fluctuated greatly (including the effect of ecological interactions of the population with its environment), such that any specific strong fitness effect is diluted as an average over time and number of experienced environments (and fitness-related phenotypes). This view would reflect the "seascape" approach to fitness landscapes \citep{Mustonen2009-cz}. A laboratory environment removes as many of these fluctuations and complexities as possible to extract the fitness (or phenotypic) effect of mutations in a well-defined environment. Thus, one main target of selection in the laboratory may be to streamline molecular processes and to remove all unnecessary mechanisms allowing an organism to respond to environmental fluctuations, e.g., by knocking out genes with premature stop codons \citep[see, e.g.][]{Kvitek2013-jz,Kryazhimskiy2014-yn}. In addition, fluctuations could result both from a changing environment altering the fitness of genotypes and from population size changes altering the strength of genetic drift, thereby making fitness differences "invisible". Just as \citet{Wright1932-ds} argued, population bottlenecks could effectively smoothen the shape of the fitness landscape perceived by an evolving population in nature.

%The same arguments apply to the reasoning for why epistasis is pervasive in fitness landscape studies, yet it is often thought to be negligible in the study of evolution in nature. 
In the extreme case, all but the strongest epistasis could become invisible to natural selection and the fitness landscape could effectively smoothen to an almost flat surface with holes of strongly deleterious, incompatible genotypes, as described in Gavrilets' holey fitness landscape model \citep{Gavrilets2004-bm}. Relatedly, the study of combinatorially complete fitness landscapes might be misleading our intuition about evolution on these landscapes. The epistasis experienced by an evolving population based on its segregating variation is likely different from the observed epistasis when considering the entire possible fitness landscape \citep{Whitlock1995-zo}. Quantifying these differences and finding measures of ``realized epistasis" is a challenge because the segregating variation strongly depends on evolutionary parameters such as ploidy, recombination, population size and mutation rate.
\subsection{At which level do ecology and the environment modify fitness landscapes, and which traces does that leave?}
\begin{figure}[ht]
\includegraphics[width=4in]{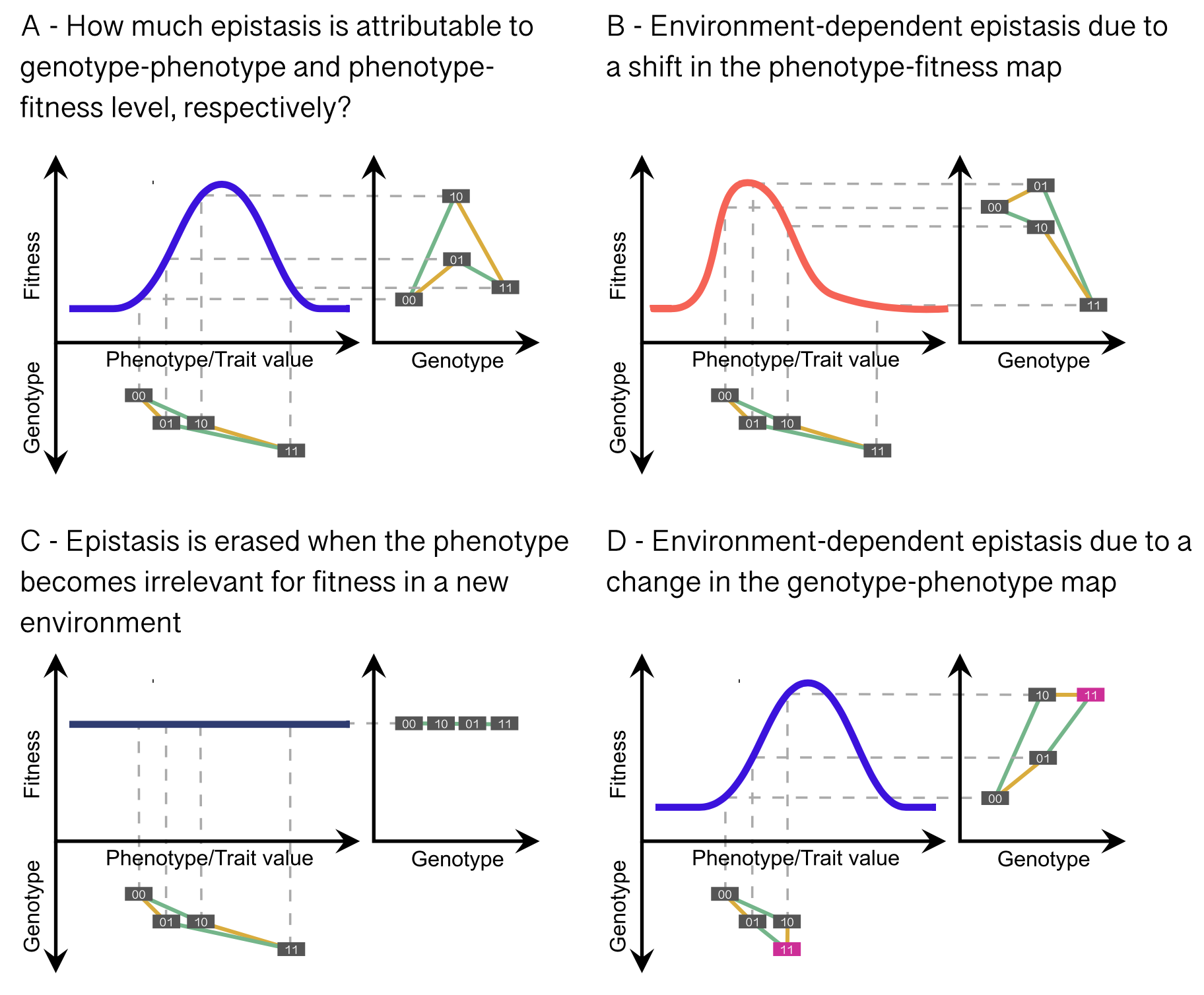}
\caption{Illustration of challenges facing the dissection of genotype-fitness relationships and the study of environment-dependent epistasis. (A) Epistasis on the genotype-phenotype level interacts with non-linear phenotype-fitness relationships. (B) A shift in the phenotypic optimum introduces environment-dependent epistasis. (C) Epistasis for fitness is erased when a trait becomes irrelevant for fitness in a new environment. (D) Epigenetic modification can introduce a change in the genotype-phenotype and genotype-fitness maps (indicated by the pink genotype) without affecting the phenotype-fitness map. Each panel represents a simplified sketch of genotype-phenotype-fitness maps. The upper left quadrant shows a phenotype-fitness map (colored curve) in a one-dimensional phenotype space. The genotype-phenotype level is represented by two mutations (yellow and green lines) interacting epistatically (lower left quadrant; positive epistasis in panels (A)-(C), negative epistasis in panel (D)). Binary genotypes are indicated by gray boxes, connected by single-step mutations (yellow and green lines).}
\label{fig4}
\end{figure}

It is unknown which part of the genotype-phenotype-fitness relationship is most affected by changes in the environment, and thus, what causes changes in epistasis across environments (see \textbf{Figure \ref{fig4}B-D}). Empirical and theoretical studies could address this question by developing models allowing for environment-tunable epistasis on the genotype-phenotype and the phenotype-fitness levels, and by designing approaches to extract those respective pieces of the fitness landscape. As \textbf{Figure \ref{fig4}B-C} show, a change in the phenotype-fitness mapping straightforwardly leads to changes in the genotype-fitness landscape and thus, changes in epistasis across environments. This can, for example, occur when a previously fitness-relevant phenotype becomes irrelevant in a different environment (e.g., camouflage in a dark or turbid environment, fast swimming in the absence of a predator, or antibiotic resistance in an antibiotic-free environment; \textbf{Figure \ref{fig4}C}). On the other hand, the environment can change epistasis in the genotype-phenotype map through environmentally induced (potentially transgenerational) epigenetic modification without changing the phenotype-fitness map, for example by silencing a gene involved in an epistatic interaction (\textbf{Figure \ref{fig4}D}).
Epistasis constrains evolutionary trajectories. If epistasis changes across environments, this implies that genomes in the present may contain information on the environment(s) the population has experienced in the past. Thus, the inference of evolutionary history from genomic samples may be linked to (or confounded by) the ecological history of a population \citep{Das2021-kj}. Whether and in which conditions this concept is applicable to more general fitness landscapes, and whether it can be translated into an inference framework for genomic data, is currently unknown. Moreover, it is unknown how the contingency of evolutionary trajectories caused by epistasis (in stable or fluctuating environments), and the frequency-dependent selection caused by epistatic selection in admixed populations, may confound population-genetic inference of evolutionary histories or selection. As a first step and proof-of-concept to address this question, features of probabilistic fitness landscapes with widespread epistasis could be incorporated into genomic simulations to quantify how the presence of epistasis changes subsequent population-genetic inference.

Finally, epistasis can span all biological levels of organization: within or between genes, between host and symbiont, and possibly between species (see, e.g., \citealp{Sorensen2021-xj, Buskirk2020-kt}). Incorporating these levels into a theory of co-evolutionary fitness landscape is an exciting challenge. One theoretical approach could be to concatenate the genomes of interacting species into one ``meta-genome" as a node in the fitness landscape -- but how would such a model reflect the potentially different generation times and population sizes of the interacting species? Also within species, competition for resources could lead to a dynamic fitness landscape in which the fitness of one genotype in a population depends on the frequency of the others. For example, epistasis on the genotype-trait level is likely to affect the long-term diversity of the population as it propagates into the (frequency-dependent) genotype-fitness map.
\section{FINAL REMARKS}
Epistasis and fitness landscapes are an active subject of study across different fields of biology. In evolutionary biology, this subject stands a bit isolated, if not ``esoteric", as the role of epistasis in evolution is often believed to be marginal except as a mechanism to maintain species integrity. It is time to acknowledge the prevalence of epistasis, which has left its marks in contingent evolutionary trajectories. New theoretical and empirical evidence of the extent and architecture of epistasis, and new possibilities to study complex interactions theoretically and empirically, call for a reevaluation of how epistasis affects (and has affected) evolution and whether and how epistasis might confound inferences from evolutionary studies. Moreover, evolutionary perspectives on genotype-phenotype-fitness relationships and the importance of environment-dependent epistasis are essential to further develop the biological relevance of fitness landscape studies. Bridging the gap between theoretical and experimental fitness landscapes and the role of epistasis in the wild requires interdisciplinary research integrating the viewpoints of molecular biology, systems biology, and ecology and evolution.
% Summary Points
\begin{summary}[SUMMARY POINTS]
\begin{enumerate}
\item Fitness landscape theory shows how epistasis constrains adaptation and promotes divergence, and indicates how experimental choices influence the expected epistasis in an empirical fitness landscape.
\item Epistasis is prevalent in empirical fitness landscapes, with a common pattern of diminishing-returns epistasis, where the beneficial effect of a substitution is smaller when it appears in a fitter genetic background. 
\item There is increasing evidence that epistasis can be environment-dependent. Theory indicates that genomes evolved in a fitness landscape with environment-dependent epistasis may carry information on their ``ecological history".
\end{enumerate}
\end{summary}
% Future Issues
\begin{issues}[FUTURE ISSUES]
\begin{enumerate}
\item How can the existing fitness landscape theory be applied, generalized or extended to cases of diploid/recombining/polymorphic/small populations that evolve in changing environments and in the presence of ecological interactions?
\item How much does the prevalence of epistasis and the resulting historical contingency affect evolutionary inference?
\item How frequently and in which way does the environment alter epistasis on the genotype-phenotype level and the phenotype-fitness level, respectively? 
\end{enumerate}
\end{issues}
%Disclosure
\section*{DISCLOSURE STATEMENT}
The author is not aware of any affiliations, memberships, funding, or financial holdings that might be perceived as affecting the objectivity of this review. 
% Acknowledgements
\section*{ACKNOWLEDGMENTS}
I am grateful for the thoughtful comments and suggestions from Kathleen Donohue, Sergey Kryazhimskiy, Joachim Krug, Christopher Martin, Austin Patton, Thomas Bataillon, Marco Louro, Adamandia Kapopoulou and David McLeod, for discussion and inspiration provided by the THEE lab, for precious support by Vitor Sousa and Chaitanya Gokhale, and funding from ERC Starting Grant 804569 (FIT2GO) and HFSP Young Investigator Grant RGY0081/2020. 
% References
\bibliography{library.bib}

\begin{thebibliography}{}
\expandafter\ifx\csname natexlab\endcsname\relax\def\natexlab#1{#1}\fi

\bibitem[Adams et~al.(2019)Adams, Kinney, Walczak \& Mora]{Adams2019-vg}
Adams RM, Kinney JB, Walczak AM, Mora T. 2019.
Epistasis in a fitness landscape defined by {Antibody-Antigen} binding free
  energy.
\textit{Cell Syst} 8(1):86--93.e3

\bibitem[Agarwala \& Fisher(2019)]{Agarwala2019-mj}
Agarwala A, Fisher DS. 2019.
Adaptive walks on high-dimensional fitness landscapes and seascapes with
  distance-dependent statistics.
\textit{Theor. Popul. Biol.} 130:13--49

\bibitem[Aguilar-Rodr{\'\i}guez et~al.(2017)Aguilar-Rodr{\'\i}guez, Payne \&
  Wagner]{Aguilar-Rodriguez2017-gz}
Aguilar-Rodr{\'\i}guez J, Payne JL, Wagner A. 2017.
A thousand empirical adaptive landscapes and their navigability.
\textit{Nat Ecol Evol} 1(2):45

\bibitem[Aita et~al.(2001)Aita, Iwakura \& Husimi]{Aita2001-ub}
Aita T, Iwakura M, Husimi Y. 2001.
A cross-section of the fitness landscape of dihydrofolate reductase.
\textit{Protein Eng.} 14(9):633--638

\bibitem[Aita et~al.(2000)Aita, Uchiyama, Inaoka, Nakajima, Kokubo \&
  Husimi]{Aita2000-qw}
Aita T, Uchiyama H, Inaoka T, Nakajima M, Kokubo T, Husimi Y. 2000.
Analysis of a local fitness landscape with a model of the rough mt. fuji-type
  landscape: application to prolyl endopeptidase and thermolysin.
\textit{Biopolymers} 54(1):64--79

\bibitem[Arnegard et~al.(2014)Arnegard, McGee, Matthews, Marchinko, Conte
  et~al.]{Arnegard2014-pu}
Arnegard ME, McGee MD, Matthews B, Marchinko KB, Conte GL, et~al. 2014.
Genetics of ecological divergence during speciation.
\textit{Nature} 511(7509):307--311

\bibitem[Athavale et~al.(2014)Athavale, Spicer \& Chen]{Athavale2014-bk}
Athavale SS, Spicer B, Chen IA. 2014.
Experimental fitness landscapes to understand the molecular evolution of
  {RNA-based} life.
\textit{Curr. Opin. Chem. Biol.} 22:35--39

\bibitem[Bakerlee et~al.(2022)Bakerlee, Nguyen~Ba, Shulgina, Rojas~Echenique \&
  Desai]{Bakerlee2022-bc}
Bakerlee CW, Nguyen~Ba AN, Shulgina Y, Rojas~Echenique JI, Desai MM. 2022.
Idiosyncratic epistasis leads to global fitness-correlated trends

\bibitem[Bank et~al.(2012)Bank, B{\"u}rger \& Hermisson]{Bank2012-ao}
Bank C, B{\"u}rger R, Hermisson J. 2012.
The limits to parapatric speciation: {Dobzhansky--Muller} incompatibilities in
  a {Continent--Island} model.
\textit{Genetics} 191(3):845--863

\bibitem[Bank et~al.(2015)Bank, Hietpas, Jensen \& Bolon]{Bank2015-kh}
Bank C, Hietpas RT, Jensen JD, Bolon DNA. 2015.
A systematic survey of an intragenic epistatic landscape.
\textit{Mol. Biol. Evol.} 32(1):229--238

\bibitem[Bank et~al.(2014)Bank, Hietpas, Wong, Bolon \& Jensen]{Bank2014-xw}
Bank C, Hietpas RT, Wong A, Bolon DN, Jensen JD. 2014.
A bayesian {MCMC} approach to assess the complete distribution of fitness
  effects of new mutations: uncovering the potential for adaptive walks in
  challenging environments.
\textit{Genetics} 196(3):841--852

\bibitem[Bank et~al.(2016)Bank, Matuszewski, Hietpas \& Jensen]{Bank2016-fn}
Bank C, Matuszewski S, Hietpas RT, Jensen JD. 2016.
On the (un)predictability of a large intragenic fitness landscape.
\textit{Proc. Natl. Acad. Sci. U. S. A.} 113(49):14085--14090

\bibitem[Barton \& Bengtsson(1986)]{Barton1986-bi}
Barton N, Bengtsson BO. 1986.
The barrier to genetic exchange between hybridising populations.
\textit{Heredity} 57 ( Pt 3):357--376

\bibitem[Bendixsen et~al.(2021)Bendixsen, Pollock, Peri \&
  Hayden]{Bendixsen2021-iz}
Bendixsen DP, Pollock TB, Peri G, Hayden EJ. 2021.
Experimental resurrection of ancestral mammalian {CPEB3} ribozymes reveals deep
  functional conservation.
\textit{Mol. Biol. Evol.} 38(7):2843--2853

\bibitem[Blanckaert \& Bank(2018)]{Blanckaert2018-ku}
Blanckaert A, Bank C. 2018.
In search of the goldilocks zone for hybrid speciation.
\textit{PLoS Genet.} 14(9):e1007613

\bibitem[Blanckaert et~al.(2020)Blanckaert, Bank \&
  Hermisson]{Blanckaert2020-yb}
Blanckaert A, Bank C, Hermisson J. 2020.
The limits to parapatric speciation 3: evolution of strong reproductive
  isolation in presence of gene flow despite limited ecological
  differentiation.
\textit{Philos. Trans. R. Soc. Lond. B Biol. Sci.} 375(1806):20190532

\bibitem[Blanquart et~al.(2014)Blanquart, Achaz, Bataillon \&
  Tenaillon]{Blanquart2014-jj}
Blanquart F, Achaz G, Bataillon T, Tenaillon O. 2014.
Properties of selected mutations and genotypic landscapes under fisher's
  geometric model.
\textit{Evolution} 68(12):3537--3554

\bibitem[Bordbar et~al.(2014)Bordbar, Monk, King \& Palsson]{Bordbar2014-cg}
Bordbar A, Monk JM, King ZA, Palsson BO. 2014.
Constraint-based models predict metabolic and associated cellular functions.
\textit{Nat. Rev. Genet.} 15(2):107--120

\bibitem[Buskirk et~al.(2020)Buskirk, Rokes \& Lang]{Buskirk2020-kt}
Buskirk SW, Rokes AB, Lang GI. 2020.
Adaptive evolution of nontransitive fitness in yeast.
\textit{Elife} 9:e62238

\bibitem[Chevin(2011)]{Chevin2011-zu}
Chevin LM. 2011.
On measuring selection in experimental evolution.
\textit{Biol. Lett.} 7(2):210--213

\bibitem[Concepci{\'o}n-Acevedo et~al.(2015)Concepci{\'o}n-Acevedo, Weiss,
  Chaudhry \& Levin]{Concepcion-Acevedo2015-nn}
Concepci{\'o}n-Acevedo J, Weiss HN, Chaudhry WN, Levin BR. 2015.
Malthusian parameters as estimators of the fitness of microbes: A cautionary
  tale about the low side of high throughput.
\textit{PLoS One} 10(6):e0126915

\bibitem[Corbett-Detig et~al.(2013)Corbett-Detig, Zhou, Clark, Hartl \&
  Ayroles]{Corbett-Detig2013-rr}
Corbett-Detig RB, Zhou J, Clark AG, Hartl DL, Ayroles JF. 2013.
Genetic incompatibilities are widespread within species.
\textit{Nature} 504(7478):135--137

\bibitem[Costanzo et~al.(2010)Costanzo, Baryshnikova, Bellay, Kim, Spear
  et~al.]{Costanzo2010-cn}
Costanzo M, Baryshnikova A, Bellay J, Kim Y, Spear ED, et~al. 2010.
The genetic landscape of a cell.
\textit{Science} 327(5964):425--431

\bibitem[Costanzo et~al.(2019)Costanzo, Kuzmin, van Leeuwen, Mair, Moffat
  et~al.]{Costanzo2019-rb}
Costanzo M, Kuzmin E, van Leeuwen J, Mair B, Moffat J, et~al. 2019.
Global genetic networks and the {Genotype-to-Phenotype} relationship.
\textit{Cell} 177(1):85--100

\bibitem[Cote-Hammarlof et~al.(2021)Cote-Hammarlof, Fragata, Flynn, Mavor,
  Zeldovich et~al.]{Cote-Hammarlof2021-ow}
Cote-Hammarlof PA, Fragata I, Flynn J, Mavor D, Zeldovich KB, et~al. 2021.
The adaptive potential of the middle domain of yeast hsp90.
\textit{Mol. Biol. Evol.} 38(2):368--379

\bibitem[Coughlan \& Matute(2020)]{Coughlan2020-ik}
Coughlan JM, Matute DR. 2020.
The importance of intrinsic postzygotic barriers throughout the speciation
  process.
\textit{Philos. Trans. R. Soc. Lond. B Biol. Sci.} 375(1806):20190533

\bibitem[Das et~al.(2020)Das, Direito, Waclaw, Allen \& Krug]{Das2020-xt}
Das SG, Direito SO, Waclaw B, Allen RJ, Krug J. 2020.
Predictable properties of fitness landscapes induced by adaptational tradeoffs.
\textit{Elife} 9:e55155

\bibitem[Das et~al.(2021)Das, Krug \& Mungan]{Das2021-kj}
Das SG, Krug J, Mungan M. 2021.
A driven disordered systems approach to biological evolution in changing
  environments.
\textit{bioRxiv} (2108.06170):10.1101/2021.08.13.456229

\bibitem[de~Visser \& Krug(2014)]{De_Visser2014-vr}
de~Visser JAGM, Krug J. 2014.
Empirical fitness landscapes and the predictability of evolution.
\textit{Nat. Rev. Genet.} 15(7):480--490

\bibitem[Domingo et~al.(2019)Domingo, Baeza-Centurion \&
  Lehner]{Domingo2019-vy}
Domingo J, Baeza-Centurion P, Lehner B. 2019.
The causes and consequences of genetic interactions (epistasis).
\textit{Annu. Rev. Genomics Hum. Genet.} 20:433--460

\bibitem[Draghi \& Plotkin(2013)]{Draghi2013-hl}
Draghi JA, Plotkin JB. 2013.
Selection biases the prevalence and type of epistasis along adaptive
  trajectories.
\textit{Evolution} 67(11):3120--3131

\bibitem[Ferretti et~al.(2016)Ferretti, Schmiegelt, Weinreich, Yamauchi,
  Kobayashi et~al.]{Ferretti2016-ss}
Ferretti L, Schmiegelt B, Weinreich D, Yamauchi A, Kobayashi Y, et~al. 2016.
Measuring epistasis in fitness landscapes: The correlation of fitness effects
  of mutations.
\textit{J. Theor. Biol.} 396:132--143

\bibitem[Fisher(1930)]{Fisher1930-mk}
Fisher RA. 1930.
The genetical theory of natural selection.
Clarendon Press

\bibitem[Flynn et~al.(2020)Flynn, Rossouw, Cote-Hammarlof, Fragata, Mavor
  et~al.]{Flynn2020-si}
Flynn JM, Rossouw A, Cote-Hammarlof P, Fragata I, Mavor D, et~al. 2020.
Comprehensive fitness maps of hsp90 show widespread environmental dependence.
\textit{Elife} 9

\bibitem[Flynn et~al.(2013)Flynn, Cooper, Moore \& Cooper]{Flynn2013-gf}
Flynn KM, Cooper TF, Moore FBG, Cooper VS. 2013.
The environment affects epistatic interactions to alter the topology of an
  empirical fitness landscape.
\textit{PLoS Genet.} 9(4):e1003426

\bibitem[Fowler \& Fields(2014)]{Fowler2014-qr}
Fowler DM, Fields S. 2014.
Deep mutational scanning: a new style of protein science.
\textit{Nat. Methods} 11(8):801--807

\bibitem[Fragata et~al.(2019)Fragata, Blanckaert, Dias~Louro, Liberles \&
  Bank]{Fragata2019-tn}
Fragata I, Blanckaert A, Dias~Louro MA, Liberles DA, Bank C. 2019.
Evolution in the light of fitness landscape theory.
\textit{Trends Ecol. Evol.} 34(1):69--82

\bibitem[Fra{\"\i}sse et~al.(2016)Fra{\"\i}sse, Gunnarsson, Roze, Bierne \&
  Welch]{Fraisse2016-pp}
Fra{\"\i}sse C, Gunnarsson PA, Roze D, Bierne N, Welch JJ. 2016.
The genetics of speciation: Insights from fisher's geometric model.
\textit{Evolution} 70(7):1450--1464

\bibitem[Gavrilets(1997)]{Gavrilets1997-ed}
Gavrilets S. 1997.
Hybrid zones with dobzhansky-type epistatic selection.
\textit{Evolution} 51(4):1027--1035

\bibitem[Gavrilets(2004)]{Gavrilets2004-bm}
Gavrilets S. 2004.
Fitness landscapes and the origin of species.
Princeton University Press

\bibitem[Gillespie(1984)]{Gillespie1984-nl}
Gillespie JH. 1984.
Molecular evolution over the mutational landscape.
\textit{Evolution} 38(5):1116--1129

\bibitem[Gong \& Bloom(2014)]{Gong2014-zb}
Gong LI, Bloom JD. 2014.
Epistatically interacting substitutions are enriched during adaptive protein
  evolution.
\textit{PLoS Genet.} 10(5):e1004328

\bibitem[Gong et~al.(2013)Gong, Suchard \& Bloom]{Gong2013-wm}
Gong LI, Suchard MA, Bloom JD. 2013.
Stability-mediated epistasis constrains the evolution of an influenza protein.
\textit{Elife} 2:e00631

\bibitem[Gorter et~al.(2018)Gorter, Aarts, Zwaan \& de~Visser]{Gorter2018-zq}
Gorter FA, Aarts MGM, Zwaan BJ, de~Visser JAGM. 2018.
Local fitness landscapes predict yeast evolutionary dynamics in directionally
  changing environments.
\textit{Genetics} 208(1):307--322

\bibitem[Goutelle et~al.(2008)Goutelle, Maurin, Rougier, Barbaut, Bourguignon
  et~al.]{Goutelle2008-yt}
Goutelle S, Maurin M, Rougier F, Barbaut X, Bourguignon L, et~al. 2008.
The hill equation: a review of its capabilities in pharmacological modelling.
\textit{Fundam. Clin. Pharmacol.} 22(6):633--648

\bibitem[Greene \& Crona(2014)]{Greene2014-uu}
Greene D, Crona K. 2014.
The changing geometry of a fitness landscape along an adaptive walk.
\textit{PLoS Comput. Biol.} 10(5):e1003520

\bibitem[Gros et~al.(2009)Gros, Le~Nagard \& Tenaillon]{Gros2009-kq}
Gros PA, Le~Nagard H, Tenaillon O. 2009.
The evolution of epistasis and its links with genetic robustness, complexity
  and drift in a phenotypic model of adaptation.
\textit{Genetics} 182(1):277--293

\bibitem[Hall et~al.(2019)Hall, Karkare, Cooper, Bank, Cooper \&
  Moore]{Hall2019-pp}
Hall AE, Karkare K, Cooper VS, Bank C, Cooper TF, Moore FBG. 2019.
Environment changes epistasis to alter trade-offs along alternative
  evolutionary paths.
\textit{Evolution} 73(10):2094--2105

\bibitem[Hansen \& P{\'e}labon(2021)]{Hansen2021-yl}
Hansen TF, P{\'e}labon C. 2021.
Evolvability: A {Quantitative-Genetics} perspective.
\textit{Annu. Rev. Ecol. Evol. Syst.} 52(1):153--175

\bibitem[Harmand et~al.(2017)Harmand, Gallet, Jabbour-Zahab, Martin \&
  Lenormand]{Harmand2017-ic}
Harmand N, Gallet R, Jabbour-Zahab R, Martin G, Lenormand T. 2017.
Fisher's geometrical model and the mutational patterns of antibiotic resistance
  across dose gradients.
\textit{Evolution} 71(1):23--37

\bibitem[Hietpas et~al.(2012)Hietpas, Roscoe, Jiang \& Bolon]{Hietpas2012-iw}
Hietpas R, Roscoe B, Jiang L, Bolon DNA. 2012.
Fitness analyses of all possible point mutations for regions of genes in yeast.
\textit{Nat. Protoc.} 7(7):1382--1396

\bibitem[Hurst et~al.(2004)Hurst, P{\'a}l \& Lercher]{Hurst2004-vn}
Hurst LD, P{\'a}l C, Lercher MJ. 2004.
The evolutionary dynamics of eukaryotic gene order.
\textit{Nat. Rev. Genet.} 5(4):299--310

\bibitem[Hwang et~al.(2017)Hwang, Park \& Krug]{Hwang2017-uu}
Hwang S, Park SC, Krug J. 2017.
Genotypic complexity of fisher's geometric model.
\textit{Genetics} 206(2):1049--1079

\bibitem[Hwang et~al.(2018)Hwang, Schmiegelt, Ferretti \& Krug]{Hwang2018-an}
Hwang S, Schmiegelt B, Ferretti L, Krug J. 2018.
Universality classes of interaction structures for {NK} fitness landscapes.
\textit{J. Stat. Phys.} 172(1):226--278

\bibitem[Johnson et~al.(2019)Johnson, Martsul, Kryazhimskiy \&
  Desai]{Johnson2019-ya}
Johnson MS, Martsul A, Kryazhimskiy S, Desai MM. 2019.
Higher-fitness yeast genotypes are less robust to deleterious mutations.
\textit{Science} 366(6464):490--493

\bibitem[Karageorgi et~al.(2019)Karageorgi, Groen, Sumbul, Pelaez, Verster
  et~al.]{Karageorgi2019-os}
Karageorgi M, Groen SC, Sumbul F, Pelaez JN, Verster KI, et~al. 2019.
Genome editing retraces the evolution of toxin resistance in the monarch
  butterfly.
\textit{Nature} 574(7778):409--412

\bibitem[Kauffman \& Levin(1987)]{Kauffman1987-ed}
Kauffman S, Levin S. 1987.
Towards a general theory of adaptive walks on rugged landscapes.
\textit{J. Theor. Biol.} 128(1):11--45

\bibitem[Kauffman \& Weinberger(1989)]{Kauffman1989-wb}
Kauffman SA, Weinberger ED. 1989.
The {NK} model of rugged fitness landscapes and its application to maturation
  of the immune response.
\textit{J. Theor. Biol.} 141(2):211--245

\bibitem[Kemble et~al.(2020)Kemble, Eisenhauer, Couce, Chapron, Magnan
  et~al.]{Kemble2020-kb}
Kemble H, Eisenhauer C, Couce A, Chapron A, Magnan M, et~al. 2020.
Flux, toxicity, and expression costs generate complex genetic interactions in a
  metabolic pathway.
\textit{Sci Adv} 6(23):eabb2236

\bibitem[Khan et~al.(2011)Khan, Dinh, Schneider, Lenski \& Cooper]{Khan2011-lr}
Khan AI, Dinh DM, Schneider D, Lenski RE, Cooper TF. 2011.
Negative epistasis between beneficial mutations in an evolving bacterial
  population.
\textit{Science} 332(6034):1193--1196

\bibitem[King et~al.(2016)King, Lu, Dr{\"a}ger, Miller, Federowicz
  et~al.]{King2016-ld}
King ZA, Lu J, Dr{\"a}ger A, Miller P, Federowicz S, et~al. 2016.
{BiGG} models: A platform for integrating, standardizing and sharing
  genome-scale models.
\textit{Nucleic Acids Res.} 44(D1):D515--22

\bibitem[Kinsler et~al.(2020)Kinsler, Geiler-Samerotte \&
  Petrov]{Kinsler2020-rz}
Kinsler G, Geiler-Samerotte K, Petrov DA. 2020.
Fitness variation across subtle environmental perturbations reveals local
  modularity and global pleiotropy of adaptation.
\textit{Elife} 9:e61271

\bibitem[Krug(2021)]{Krug2021-dp}
Krug J. 2021.
Accessibility percolation in random fitness landscapes. In
  \textit{Probabilistic Structures in Evolution}, ed. Wa~Baake~E. EMS Press,
  Berlin,  1--22

\bibitem[Kryazhimskiy(2021)]{Kryazhimskiy2021-sv}
Kryazhimskiy S. 2021.
Emergence and propagation of epistasis in metabolic networks.
\textit{Elife} 10:e60200

\bibitem[Kryazhimskiy et~al.(2014)Kryazhimskiy, Rice, Jerison \&
  Desai]{Kryazhimskiy2014-yn}
Kryazhimskiy S, Rice DP, Jerison ER, Desai MM. 2014.
Global epistasis makes adaptation predictable despite sequence-level
  stochasticity.
\textit{Science} 344(6191):1519--1522

\bibitem[Kvitek \& Sherlock(2013)]{Kvitek2013-jz}
Kvitek DJ, Sherlock G. 2013.
Whole genome, whole population sequencing reveals that loss of signaling
  networks is the major adaptive strategy in a constant environment.
\textit{PLoS Genet.} 9(11):e1003972

\bibitem[L{\"a}ssig et~al.(2017)L{\"a}ssig, Mustonen \& Walczak]{Lassig2017-vw}
L{\"a}ssig M, Mustonen V, Walczak AM. 2017.
Predicting evolution.
\textit{Nat Ecol Evol} 1(3):77

\bibitem[Li \& Zhang(2018)]{Li2018-gs}
Li C, Zhang J. 2018.
Multi-environment fitness landscapes of a {tRNA} gene.
\textit{Nat Ecol Evol} 2(6):1025--1032

\bibitem[Lindsey et~al.(2013)Lindsey, Gallie, Taylor \& Kerr]{Lindsey2013-fd}
Lindsey HA, Gallie J, Taylor S, Kerr B. 2013.
Evolutionary rescue from extinction is contingent on a lower rate of
  environmental change.
\textit{Nature} 494(7438):463--467

\bibitem[Lyons et~al.(2020)Lyons, Zou, Xu \& Zhang]{Lyons2020-lf}
Lyons DM, Zou Z, Xu H, Zhang J. 2020.
Idiosyncratic epistasis creates universals in mutational effects and
  evolutionary trajectories.
\textit{Nat Ecol Evol} 4(12):1685--1693

\bibitem[Mackay(2013)]{Mackay2013-tp}
Mackay TFC. 2013.
Epistasis and quantitative traits: using model organisms to study gene--gene
  interactions.
\textit{Nat. Rev. Genet.} 15(1):22--33

\bibitem[Marcusson et~al.(2009)Marcusson, Frimodt-M{\o}ller \&
  Hughes]{Marcusson2009-nf}
Marcusson LL, Frimodt-M{\o}ller N, Hughes D. 2009.
Interplay in the selection of fluoroquinolone resistance and bacterial fitness.
\textit{PLoS Pathog.} 5(8):e1000541

\bibitem[Marmier et~al.(2019)Marmier, Weigt \& Bitbol]{Marmier2019-db}
Marmier G, Weigt M, Bitbol AF. 2019.
Phylogenetic correlations can suffice to infer protein partners from sequences.
\textit{PLoS Comput. Biol.} 15(10):e1007179

\bibitem[Martin(2016)]{Martin2016-vp}
Martin CH. 2016.
Context dependence in complex adaptive landscapes: frequency and
  trait-dependent selection surfaces within an adaptive radiation of caribbean
  pupfishes

\bibitem[Martin \& Gould(2020)]{Martin2020-ws}
Martin CH, Gould KJ. 2020.
Surprising spatiotemporal stability of a multi-peak fitness landscape revealed
  by independent field experiments measuring hybrid fitness.
\textit{Evol Lett} 4(6):530--544

\bibitem[Martin \& Wainwright(2013)]{Martin2013-jz}
Martin CH, Wainwright PC. 2013.
Multiple fitness peaks on the adaptive landscape drive adaptive radiation in
  the wild.
\textit{Science} 339(6116):208--211

\bibitem[Martin et~al.(2007)Martin, Elena \& Lenormand]{Martin2007-mq}
Martin G, Elena SF, Lenormand T. 2007.
Distributions of epistasis in microbes fit predictions from a fitness landscape
  model.
\textit{Nat. Genet.} 39(4):555--560

\bibitem[Martin \& Lenormand(2006)]{Martin2006-ad}
Martin G, Lenormand T. 2006.
The fitness effect of mutations across environments: a survey in light of
  fitness landscape models.
\textit{Evolution} 60(12):2413--2427

\bibitem[Martin \& Lenormand(2015)]{Martin2015-qz}
Martin G, Lenormand T. 2015.
The fitness effect of mutations across environments: Fisher's geometrical model
  with multiple optima.
\textit{Evolution} 69(6):1433--1447

\bibitem[Matuszewski et~al.(2014)Matuszewski, Hermisson \&
  Kopp]{Matuszewski2014-ot}
Matuszewski S, Hermisson J, Kopp M. 2014.
Fisher's geometric model with a moving optimum.
\textit{Evolution} 68(9):2571--2588

\bibitem[McCandlish(2011)]{McCandlish2011-ve}
McCandlish DM. 2011.
Visualizing fitness landscapes.
\textit{Evolution} 65(6):1544--1558

\bibitem[Moradigaravand \& Engelst{\"a}dter(2012)]{Moradigaravand2012-kw}
Moradigaravand D, Engelst{\"a}dter J. 2012.
The effect of bacterial recombination on adaptation on fitness landscapes with
  limited peak accessibility.
\textit{PLoS Comput. Biol.} 8(10):e1002735

\bibitem[Morcos et~al.(2011)Morcos, Pagnani, Lunt, Bertolino, Marks
  et~al.]{Morcos2011-di}
Morcos F, Pagnani A, Lunt B, Bertolino A, Marks DS, et~al. 2011.
Direct-coupling analysis of residue coevolution captures native contacts across
  many protein families.
\textit{Proc. Natl. Acad. Sci. U. S. A.} 108(49):E1293--301

\bibitem[Mustonen \& L{\"a}ssig(2009)]{Mustonen2009-cz}
Mustonen V, L{\"a}ssig M. 2009.
From fitness landscapes to seascapes: non-equilibrium dynamics of selection and
  adaptation.
\textit{Trends Genet.} 25(3):111--119

\bibitem[Neidhart et~al.(2014)Neidhart, Szendro \& Krug]{Neidhart2014-gy}
Neidhart J, Szendro IG, Krug J. 2014.
Adaptation in tunably rugged fitness landscapes: the rough mount fuji model.
\textit{Genetics} 198(2):699--721

\bibitem[Nosil et~al.(2020)Nosil, Villoutreix, de~Carvalho, Feder, Parchman \&
  Gompert]{Nosil2020-nd}
Nosil P, Villoutreix R, de~Carvalho CF, Feder JL, Parchman TL, Gompert Z. 2020.
Ecology shapes epistasis in a genotype-phenotype-fitness map for stick insect
  colour.
\textit{Nat Ecol Evol} 4(12):1673--1684

\bibitem[Nowak et~al.(2014)Nowak, Neidhart, Szendro \& Krug]{Nowak2014-wy}
Nowak S, Neidhart J, Szendro IG, Krug J. 2014.
Multidimensional epistasis and the transitory advantage of sex.
\textit{PLoS Comput. Biol.} 10(9):e1003836

\bibitem[Ono et~al.(2017)Ono, Gerstein \& Otto]{Ono2017-cz}
Ono J, Gerstein AC, Otto SP. 2017.
Widespread genetic incompatibilities between {First-Step} mutations during
  parallel adaptation of saccharomyces cerevisiae to a common environment.
\textit{PLoS Biol.} 15(1):e1002591

\bibitem[Paix{\~a}o et~al.(2015)Paix{\~a}o, Bassler \& Azevedo]{Paixao2015-zo}
Paix{\~a}o T, Bassler KE, Azevedo RBR. 2015.
Emergent speciation by multiple {Dobzhansky--Muller} incompatibilities.
\textit{bioRxiv} (doi: 10.1101/008268):10.1101/008268

\bibitem[Park et~al.(2020)Park, Hwang \& Krug]{Park2020-hn}
Park SC, Hwang S, Krug J. 2020.
Distribution of the number of fitness maxima in fisher's geometric model.
\textit{J. Phys. A: Math. Theor.} 53(38):385601

\bibitem[Patton et~al.(2021)Patton, Richards, Gould, Buie \&
  Martin]{Patton2021-mp}
Patton AH, Richards EJ, Gould KJ, Buie LK, Martin CH. 2021.
Adaptive introgression and de novo mutations increase access to novel fitness
  peaks on the fitness landscape during a vertebrate adaptive radiation.
\textit{bioRxiv} :10.1101/2021.07.01.450666

\bibitem[Phillips(2008)]{Phillips2008-tr}
Phillips PC. 2008.
Epistasis--the essential role of gene interactions in the structure and
  evolution of genetic systems.
\textit{Nat. Rev. Genet.} 9(11):855--867

\bibitem[Pinheiro et~al.(2021)Pinheiro, Warsi, Andersson \&
  L{\"a}ssig]{Pinheiro2021-qq}
Pinheiro F, Warsi O, Andersson DI, L{\"a}ssig M. 2021.
Metabolic fitness landscapes predict the evolution of antibiotic resistance.
\textit{Nat Ecol Evol} 5(5):677--687

\bibitem[Pokusaeva et~al.(2019)Pokusaeva, Usmanova, Putintseva, Espinar,
  Sarkisyan et~al.]{Pokusaeva2019-nl}
Pokusaeva VO, Usmanova DR, Putintseva EV, Espinar L, Sarkisyan KS, et~al. 2019.
An experimental assay of the interactions of amino acids from orthologous
  sequences shaping a complex fitness landscape.
\textit{PLoS Genet.} 15(4):e1008079

\bibitem[Presgraves(2003)]{Presgraves2003-yl}
Presgraves DC. 2003.
A fine-scale genetic analysis of hybrid incompatibilities in drosophila.
\textit{Genetics} 163(3):955--972

\bibitem[Reddy \& Desai(2021)]{Reddy2021-ec}
Reddy G, Desai MM. 2021.
Global epistasis emerges from a generic model of a complex trait.
\textit{Elife} 10:e64740

\bibitem[Sachdeva et~al.(2020)Sachdeva, Husain, Sheng, Wang \&
  Murugan]{Sachdeva2020-lm}
Sachdeva V, Husain K, Sheng J, Wang S, Murugan A. 2020.
Tuning environmental timescales to evolve and maintain generalists.
\textit{Proc. Natl. Acad. Sci. U. S. A.} 117(23):12693--12699

\bibitem[Sarkisyan et~al.(2016)Sarkisyan, Bolotin, Meer, Usmanova, Mishin
  et~al.]{Sarkisyan2016-sj}
Sarkisyan KS, Bolotin DA, Meer MV, Usmanova DR, Mishin AS, et~al. 2016.
Local fitness landscape of the green fluorescent protein.
\textit{Nature} 533(7603):397--401

\bibitem[Schenk et~al.(2013)Schenk, Szendro, Salverda, Krug \&
  de~Visser]{Schenk2013-tu}
Schenk MF, Szendro IG, Salverda MLM, Krug J, de~Visser JAGM. 2013.
Patterns of epistasis between beneficial mutations in an antibiotic resistance
  gene.
\textit{Mol. Biol. Evol.} 30(8):1779--1787

\bibitem[Schneemann et~al.(2020)Schneemann, De~Sanctis, Roze, Bierne \&
  Welch]{Schneemann2020-kf}
Schneemann H, De~Sanctis B, Roze D, Bierne N, Welch JJ. 2020.
The geometry and genetics of hybridization.
\textit{Evolution} 74(12):2575--2590

\bibitem[Schumer et~al.(2014)Schumer, Cui, Powell, Dresner, Rosenthal \&
  Andolfatto]{Schumer2014-da}
Schumer M, Cui R, Powell DL, Dresner R, Rosenthal GG, Andolfatto P. 2014.
High-resolution mapping reveals hundreds of genetic incompatibilities in
  hybridizing fish species.
\textit{Elife} 3:e02535

\bibitem[Seehausen et~al.(2014)Seehausen, Butlin, Keller, Wagner, Boughman
  et~al.]{Seehausen2014-rg}
Seehausen O, Butlin RK, Keller I, Wagner CE, Boughman JW, et~al. 2014.
Genomics and the origin of species.
\textit{Nat. Rev. Genet.} 15(3):176--192

\bibitem[Simon et~al.(2018)Simon, Bierne \& Welch]{Simon2018-qt}
Simon A, Bierne N, Welch JJ. 2018.
Coadapted genomes and selection on hybrids: Fisher's geometric model explains a
  variety of empirical patterns.
\textit{Evol Lett} 2(5):472--498

\bibitem[S{\o}rensen et~al.(2021)S{\o}rensen, Wood, Cameron \&
  Brockhurst]{Sorensen2021-xj}
S{\o}rensen MES, Wood AJ, Cameron DD, Brockhurst MA. 2021.
Rapid compensatory evolution can rescue low fitness symbioses following partner
  switching.
\textit{Curr. Biol.}

\bibitem[Srivastava \& Payne(2022)]{Srivastava2022-sq}
Srivastava M, Payne JL. 2022.
The transformability of genotype-phenotype landscapes

\bibitem[Storz(2018)]{Storz2018-en}
Storz JF. 2018.
Compensatory mutations and epistasis for protein function.
\textit{Curr. Opin. Struct. Biol.} 50:18--25

\bibitem[Szendro et~al.(2013)Szendro, Schenk, Franke, Krug \&
  de~Visser]{Szendro2013-pw}
Szendro IG, Schenk MF, Franke J, Krug J, de~Visser JAGM. 2013.
Quantitative analyses of empirical fitness landscapes.
\textit{J. Stat. Mech.} 2013(01):P01005

\bibitem[Tataru \& Bataillon(2020)]{Tataru2020-vv}
Tataru P, Bataillon T. 2020.
{polyDFE}: Inferring the distribution of fitness effects and properties of
  beneficial mutations from polymorphism data.
\textit{Methods Mol. Biol.} 2090:125--146

\bibitem[Tenaillon(2014)]{Tenaillon2014-ns}
Tenaillon O. 2014.
The utility of fisher's geometric model in evolutionary genetics.
\textit{Annu. Rev. Ecol. Evol. Syst.} 45:179--201

\bibitem[Trubenov{\'a} et~al.(2019)Trubenov{\'a}, Krejca, Lehre \&
  K{\"o}tzing]{Trubenova2019-xq}
Trubenov{\'a} B, Krejca MS, Lehre PK, K{\"o}tzing T. 2019.
Surfing on the seascape: Adaptation in a changing environment.
\textit{Evolution} 73(7):1356--1374

\bibitem[Turelli \& Orr(2000)]{Turelli2000-ez}
Turelli M, Orr HA. 2000.
Dominance, epistasis and the genetics of postzygotic isolation.
\textit{Genetics} 154(4):1663--1679

\bibitem[Walsh \& Lynch(2018)]{Walsh2018-ou}
Walsh B, Lynch M. 2018.
Evolution and selection of quantitative traits.
Oxford University Press

\bibitem[Wang \& Dai(2019)]{Wang2019-gb}
Wang S, Dai L. 2019.
Evolving generalists in switching rugged landscapes.
\textit{PLoS Comput. Biol.} 15(10):e1007320

\bibitem[Weinreich et~al.(2006)Weinreich, Delaney, Depristo \&
  Hartl]{Weinreich2006-dr}
Weinreich DM, Delaney NF, Depristo MA, Hartl DL. 2006.
Darwinian evolution can follow only very few mutational paths to fitter
  proteins.
\textit{Science} 312(5770):111--114

\bibitem[Whitlock et~al.(1995)Whitlock, Phillips, Moore \&
  Tonsor]{Whitlock1995-zo}
Whitlock MC, Phillips PC, Moore FBG, Tonsor SJ. 1995.
Multiple fitness peaks and epistasis.
\textit{Annu. Rev. Ecol. Syst.} 26(1):601--629

\bibitem[Wiser et~al.(2013)Wiser, Ribeck \& Lenski]{Wiser2013-uj}
Wiser MJ, Ribeck N, Lenski RE. 2013.
Long-term dynamics of adaptation in asexual populations.
\textit{Science} 342(6164):1364--1367

\bibitem[Wortel et~al.(2021)Wortel, Agashe, Bailey, Bank, Bisschop
  et~al.]{Wortel2021-ie}
Wortel MT, Agashe D, Bailey SF, Bank C, Bisschop K, et~al. 2021.
The why, what and how of predicting evolution across biology: from disease to
  biotechnology to biodiversity.
\textit{EcoEvoRxiv} :10.32942/osf.io/4u3mg

\bibitem[Wright(1931)]{Wright1931-fi}
Wright S. 1931.
Evolution in mendelian populations.
\textit{Genetics} 16(2):97--159

\bibitem[Wright(1932)]{Wright1932-ds}
Wright S. 1932.
The roles of mutation, inbreeding, crossbreeding, and selection in evolution.
\textit{Proceedings of the Sixth International Congress of Genetics} :356--366

\bibitem[Zhang(2012)]{Zhang2012-xm}
Zhang XS. 2012.
Fisher's geometrical model of fitness landscape and variance in fitness within
  a changing environment.
\textit{Evolution} 66(8):2350--2368

\end{thebibliography}

\end{document}